\RequirePackage{fix-cm}
\documentclass[smallextended]{svjour3}       
\smartqed  
\usepackage{graphicx}
\newcommand{\given}{\, \mid \,}
\usepackage[utf8x]{inputenc}
\usepackage{verbatim} 
\usepackage{multirow} 
\usepackage{amsmath}
\usepackage{multicol} 
\usepackage{graphicx}
\usepackage{amsfonts}

\usepackage{placeins}

\usepackage{hyperref}

\usepackage{subfigure}
\usepackage{float} 
\usepackage{adjustbox}
\newcommand{\E}{\mathbb{E}}
\newtheorem{thm}{Proposition}

\newcommand{\alphabar}{\overline\alpha}
\newcommand{\mubar}{\overline\mu}
\newcommand{\thetabar}{\overline\theta}

\begin{document}

\title{To snipe or not to snipe, that's the question!  
}
\subtitle{Transitions in sniping behaviour \\among competing high-frequency traders}


\author{Somayeh Kokabisaghi  
\and Andr\'e B. Dorsman \and Eric J. Pauwels 
}

\authorrunning{Kokabisaghi, Dorsman, Pauwels} 

\institute{Somayeh Kokabisaghi \at
              Centrum Wiskunde \& Informatica, Amsterdam, Netherlands.\\
              \email{karen.kokabisaghi@cwi.nl}, ORCID: 0000-0002-4589-7638
           \and
           Andr\"e B. Dorsman \at
              Vrije University, Amsterdam, Netherlands.\\\email{a.b.dorsman@vu.nl}
            \and
           Eric J. Pauwels \at
              Centrum Wiskunde \& Informatica, Amsterdam, Netherlands.\\ \email{Eric.Pauwels@cwi.nl}, 
              ORCID: 0000-0001-7518-6856
}

\date{Received: date / Accepted: date}

\maketitle

\begin{abstract}
In this paper we extend the investigation 
into the transition 
from sure to probabilistic sniping as 
introduced in Menkveld and Zoican~\cite{mz2017}. 
In that paper,  
the authors introduce a stylized version 
of a competitive game in which high frequency traders (HFTs) interact 
with each other and liquidity traders.  
The authors then  show that risk aversion plays an important role in the transition from sure to mixed (or probabilistic) sniping.  In this paper, we re-interpret and extend these conclusions  in the context of repeated games and  highlight some differences in results.  In particular, we identify  situations in which probabilistic sniping is genuinely profitable that are qualitatively different from the ones obtained in~\cite{mz2017}.
It turns out that beyond a specific 
risk aversion threshold 
the game resembles the well-known prisoner's dilemma, in that 
probabilistic sniping becomes a way to cooperate among the HFTs
that leaves all the participants better off. 
In order to turn this into a viable strategy for the 
repeated game, we show how compliance 
can be monitored through the use of  sequential statistical  testing. 


\end{abstract}

\section{Introduction}
\label{intro}

\subsection{Motivation}
\label{sct:motivation}

The burgeoning of algorithmic and high-frequency trading on modern 
 financial exchanges 
has give rise to theoretical investigations scrutinizing 
the mechanisms that underpin these markets. 
To analyse this complexity of continuous-time trading, 
a number of authors (e.g. \cite{Budish2015,mz2017}) have proposed 
stylized models of the stock market. 
Although these stylized models can not capture 
all the intricacies of the markets in detail, 
they are nonetheless very useful as they often hint 
at  interesting behaviour. As a case in point, 
Menkveld and Zoican~\cite{mz2017} designed a simple model 
(referred to as the MZ game from here on) 
to study the impact of exchange latency on liquidity. They show that faster exchanges reduce payoff risk and spread, 
but also give rise to speculative behaviours among high frequency traders (HFT). 

More specifically, they envisage a situation in which 
a group of competing but comparably equipped HFTs 
is operating alongside a large crowd of traditional 
liquidity traders (LT).   A market is created by 
selecting a single  market maker (MM) from among the HFTs. 
This market maker then publishes an ask-bid quote for 
a single financial asset on the exchange, 
and 
stands to earn the spread when this quote is hit 
by a liquidity trader.  However,  what makes this 
game interesting is that fresh news items, published at random times, push the intrinsic 
value of the  asset either up or down. 
The ensuing jump in value causes all the HFTs to race to the exchange, albeit 
it with different intentions.   The market maker attempts to cancel his own now stale 
quotes\footnote{A quote for an asset is {\bf stale} when it 
no longer reflects the most recent information.}, while the other HFTs (dubbed {\it bandits} hereafter) 
race to take advantage of the arbitrage  
opportunity created by the outdated quotes. 
This type of opportunistic behaviour 
on the part of the bandits, is called {\bf sniping}.  
The authors in 
\cite{mz2017} conclude (among other things) that increasing risk aversion in the market induces a qualitative transition in the sniping behaviour of HFTs.

While these results are intriguing  and highly non-trivial, they hinge on some 
subtle but important  features of the game.  In particular,  the authors study 
this problem in the context of a single-shot game, and focus on Nash equilibria 
as their main solution concept.  
In this paper we espouse 
an alternative view by interpreting 
the market activities of the HFTs as a {\it repeated game with an infinite 
horizon}.  
In this alternative setting a player 
can adopt more far-sighted strategies that 
attempt to maximise 
long-term gains. 
This change of viewpoint has 
a number of important consequences: 

\begin{itemize}
    \item It identifies situations in which probabilistic sniping is genuinely 
    profitable and therefore extends the 
    conclusions in \cite{mz2017} by introducing 
    qualitatively different solutions.  
    More specifically, it shows that forms 
    of informal collaboration 
    in which bandits voluntary decrease their sniping 
    frequency, results in better utility for everyone. 
    This explains why bandits are willing to engage in probabilistic sniping, 
    for which there 
    seems to be 
    no compelling reason in the original MZ model. 
    
    \item It allows us to identify two thresholds 
    (for risk aversion) that govern the transitions 
    between the three different sniping regimes:  
    from {\it sure sniping }
    to {\it probabilistic sniping} to 
    {\it no sniping}. 
    
    \item Finally, we identify a way for 
    an individual HFT to monitor the 
    behaviour of the other HFts in order to detect 
    non-compliance. This detection 
    mechanism is necessary  to deter 
    devious HFTs from exploiting trustworthy colleagues by 
    sniping more than is optimal.

\end{itemize}

\subsection{Overview and contributions of  this paper}  
\label{contributions}
The remainder of this paper is organised as follows.   

\begin{itemize}
    \item In {\bf section~\ref{sct:mz_summary}} we 
    introduce the {\bf MZ-game}, 
    and show how one can compute the utility of 
    each possible outcome.  Combining these outcomes 
    with the probabilities of the corresponding events 
    yields the expected utility for both the 
    market maker and any one of the bandits.  
    It turns out that both these utilities are linear 
     functions of the spread $s$. 
    The intersection of these two utility lines determines 
    a point of indifference $(s^*,u^*)$ 
    (see Fig.~\ref{fig:intersection_utilities}) that  
    predicts the behaviour of the agents. 
    More precisely, since at the start of the MZ game, 
    HFTs are still unsure about their role in the 
    game (market maker or bandit) they will advertise a spread $s$ that makes them indifferent 
    (in terms of expected utility) to the 
    outcome of this selection. The point of indifference 
    $(s^*,u^*)$ is therefore a focal point and 
     constitutes a Nash equilibrium of the 
    game whenever $u^*>0$.  However, this is no longer 
    the case when $u^* \leq 0$ and this is where 
    probabilistic sniping becomes relevant. 
    .

\item 
{\bf Section~\ref{sct:prob_vs_sure}} takes a detailed 
look at the concept of 
{\bf probabilistic sniping}:  rather than attempting to snipe 
at every possible opportunity ({\it sure} sniping), bandits voluntarily 
restrain themselves and only enter the 
sniping race with a probability $p<1$. Less sniping 
is obviously beneficial for the market maker but,  
under certain circumstances, this also protects bandits 
from adverse effects.   As a consequence, 
depending on the choice of the sniping probability $p$,
the point 
of 
indifference shifts to a (potentially) more favourable position 
$(s^*(p), u^*(p))$
(see Fig.~\ref{fig:prob_sniping_geom_basics}).  
In this section we characterize in detail how 
utilities (and derived quantities such as the point 
of indifference) depend on the sniping probability.

\item Whether or not probabilistic sniping results 
in better utility (for all HFTs involved) depends 
ultimately on the 
game parameters (see Fig.~\ref{fig:prob_sniping_geom}): 
changing these parameters might induce a 
{\it transition 
in optimal behaviour}, e.g. from sure sniping to 
probabilistic sniping.  
In {\bf section~\ref{sct:transition_geom}} we show how a 
simple geometric argument allows us to characterise 
the conditions under which two transitions between 
the different types of sniping occur: viz. 
from {\it sure sniping} to {\it probabilistic sniping} 
to {\it no sniping} (see sections \ref{sct:from_pure_to_prob} 
and \ref{sct:from_prob_to_non}). 
We then highlight the specific 
role of {\bf risk aversion}  
$\gamma$
in these sniping transitions and identify the relevant {\bf thresholds} $\overline\gamma_K$ and $\overline\gamma_L$ (section~\ref{sct:role_of_gamma}).  Finally, 
we introduce the optimal sniping probability 
$p^*_K$ (section~\ref{sct:p_star_K}) that results 
in the best possible utility $u^*_K$. 

\item 
Up to this point in the paper, 
the equilibria obtained under various conditions 
were optimal, but not 
necessarily Nash equilibria 
for the single-shot MZ game. 
In {\bf section~\ref{sct:app_repeated_games}} we explain what 
the implications are of shifting 
our viewpoint from 
one-shot to {\bf repeated games}.  In particular, we argue 
that probabilistic sniping will give rise 
to a new set of 
playable 
subgame-perfect Nash equilibria. 
To turn this into a workable solution, we need to 
be able to reliably detect {\it non-compliance} on the 
part of the other agents, in order to be able to 
take retaliatory measures.  We therefore quantify 
the effect that devious agents (agents that pretend to 
snipe probabilistically, but in fact snipe for sure) have 
on the utility of trustworthy agents 
(see e.g. Figs.~\ref{fig:mean_utility_fair_vs_actual_3} 
or \ref{fig:stem_utility}). 
We conclude this section 
by discussing how Wald's sequential 
probability ratio test can be used to reliably detect non-compliance (see Fig.~\ref{fig:sprt}).

\item 
Finally, in {\bf section~\ref{sct:related_work}}, 
we discuss how our work compares to related research and finish by offering 
some conclusions and suggestions for further research in 
{\bf section~\ref{sct:conclusions}}.

\item  The main part of this paper focuses on the 
conceptual framework that underpins our results. 
All the computational details 
are therefore relegated to 
sections~\ref{appx:abbrevs} 
through \ref{appx:prob_dist_unique_util} in the 
{\bf appendix}.   
Since the mathematical characterisation of the 
results often requires tedious but straightforward 
algebra, we also provide  Python notebooks 
to support all the computations.  These 
can be found: 

\begin{center}
    \framebox {\href{https://github.com/Karen-20/Karen.git}{KPD\_supplementary\_material link to GitHub}}
\end{center}

\end{itemize}

\paragraph{Contributions}

The main contributions in this paper 
can be found in sections \ref{sct:prob_vs_sure} 
through \ref{sct:app_repeated_games}, and can be summarized as follows: 

\begin{enumerate}

\item We show how the MZ model of the stock exchange can be 
re-interpreted 
as a {\it  repeated (sequential) game against nature. }
In addition, we think of the choice of sniping probability 
(for the bandits) as a choice of a pure action from 
a continuous action space, rather than a probabilistic 
mixing of two pure strategies (i.e. {\it sniping} and 
{\it non-sniping}).  
This change of viewpoint allows us to identify  
a wider set of conditions under which probabilistic 
sniping makes sense (see section~\ref{sct:prob_sniping_advantageous}) .  
Put succinctly, 
although these new equilibria are not 
Nash equilibria 
for a one-shot stage game, 
they are genuine 
subgame-perfect equilibria 
in the corresponding repeated game with 
infinite horizon. 

\item We identify 
two 
threshold values (viz. $\overline\gamma_K$ 
and $\overline\gamma_L$)
for the risk aversion factor $\gamma$ that govern 
the transition to probabilistic sniping.  
More specifically, when $\gamma$ increases beyond 
$\overline\gamma_K$,  probabilistic sniping becomes 
advantageous for all HFTs involved. Conversely, 
when $\gamma$ exceeds $\overline\gamma_L$, even 
probabilistic sniping is no longer profitable. 

\item  We show that under conditions 
for which probabilistic sniping 
is advantageous, there is an 
optimal sniping probability 
$p^*_K$ that maximises expected utilities for both 
the market maker and the bandits.  
Both the sniping probability and the corresponding 
optimal utility vary continuously as functions 
of the game parameters 
(see Fig.~\ref{fig:u_star_K_as_function_of_gamma}). 

\item When the game conditions call for 
probabilistic sniping in order to optimise utility, 
the agents are tempted to snipe for sure as this 
maximises their own utility to the detriment of 
their compliant colleagues whose utility is 
negatively affected. As a consequence, agents 
participating in such a probabilistic sniping 
regime need a strategy to detect the non-compliance 
of deceptive agents.  We show how this can 
be done using Wald's sequential probability 
ratio test (section \ref{sct:sprt}). 

\item {\bf Reproducible research: }We provide Python notebooks that will allow the 
diligent reader to check the straightforward 
but tedious calculations underpinning our 
arguments, as well as simulations that illustrate 
the fluctuations about the mean values that 
result from the theoretical approach. 

\end{enumerate}

\section{The Menkveld-Zoican (MZ) game revisited} 
\label{sct:mz_summary}

\subsection{A bird's eye view of the MZ game}
\label{sct:mz_summary_1}

As this paper builds on the  MZ paper~\cite{mz2017}, 
we briefly highlight some of the main arguments and conclusions 
from that paper. In \cite{mz2017},    
the authors define a stylized version of the 
behaviour and strategies  of 
high frequency traders (HFT) interacting with an (high-frequency) exchange.  
We defer some details to  section~\ref{sct:mz-game_ingredients} 
but, 
roughly speaking,  they envisage a game in which,  
during an initial pre-game stage, the HFTs
have to pick the value of the (half) spread $s$ 
in order to post 
a single bid-ask quote ($v \pm s$) for a financial asset 
of (commonly perceived) value $v$. 
One of the HFTs is then selected as market maker, 
whereas the others are relegated to the role of bandit. 
Only the quotes of the market maker will enter the 
order book, so he is the only one who can benefit 
from trading with liquidity traders who visit 
the exchange at random times (with rate $\mu$).  The bandits will 
try to gain some payoff by attempting to snipe 
stale quotes (see section \ref{sct:mz-game_ingredients} 
for more details). 

The game is initiated by a
trigger event which can be  either the publication of news item (changing the intrinsic value $v$ 
of the asset), or the interaction of 
a liquidity trader with 
the market maker (resulting in income for the latter). 
The latter event does not evoke any reaction from the HFTs,   
but the former alters 
the intrinsic value of the asset  and causes all the HFTs to 
race towards the exchange, albeit for different reasons. 
The market maker attempts to cancel his now stale quotes, 
whereas the bandits hope to obtain financial gain 
by sniping the 
market maker.  The outcome of this race if further 
complicated by the fact that, during the race,  another 
event (news or liquidity trade) might occur.   This is less 
likely when the exchange is fast (low exchange latency $\delta$). 

In~\cite{mz2017} the authors investigate in detail 
how the interplay between various game parameters affect 
the behaviour and strategies of the HFTs. 
Although this model is highly stylized 
and therefore somewhat unrealistic, 
it is the contention 
of the MZ authors that even under these simplifying assumptions,  
interesting and non-trivial conclusions can be drawn.  
This suggests that a more realistic model will give rise 
to even more intriguing insights.

\subsection{Primitives for the MZ game  } 
\label{sct:mz-game_ingredients}

Because they are important to understand the rest of the 
paper, we briefly recapitulate the main ingredients and assumptions governing the MZ game.  
For more details we refer to the original paper~\cite{mz2017}.
 
\begin{itemize}
\item{\bf Agents or Players}\quad  The players in this game 
is the group of $H>2$ high-frequency traders (HFTs)  who 
operate in an environment populated by an infinite number of 
(uninformed) 
liquidity traders (LT). All HFTs have simultaneous and instantaneous 
access to all public information affecting the market. At the start 
of the game, one HFT is assigned to the 
role of market maker.  The remaining ($H-1$) HFTs take up the role of  
 high-frequency bandits (HFB), intent on financial gain 
by sniping the stale quotes whenever an opportunity presents itself.  
A final important characteristic of all HFTs is that they 
are {\it risk-averse}, i.e. 
the utility of
{\it negative} pay-offs is inflated by a factor $\gamma >1$.

\item{\bf Exchange latency $\delta$}    \quad 
measures the time delay $\delta$ between 
the trigger event (that sets off the game) and the arrival 
of one HFT 
at the matching engine
where each order is actually processed. 
This delay determines the time span over which the race 
game is played out and,  as such, has a direct impact on the 
expected number of additional events that might occur 
during the race. 

\item {\bf Exogenous events}  The behaviour of the HFTs 
is affected by two types of exogenous events that happen independent 
of the HFTs actions: 
\begin{enumerate}
    \item {\bf Publication of news item  causing a common value shock: }  the publications of news (either good or bad)  is observed by everyone, and changes the 
    intrinsic value of an asset. This process is modelled 
    as a 
    Poisson process with rate $\alpha$ (number of 
    news events per 
    time unit). The size of the shock 
    is fixed at $\pm \sigma$. 
    Throughout this paper we are assuming that 
    the (half) spread ($s$) is (strictly) less 
    than $\sigma$ ($0 < s< \sigma$).
    \item {\bf Private value shock:} These 
    are only known to the LTs and cause 
    one of them 
    them to join 
    the matcher queue
    to hit the outstanding quote (resulting 
    in income for the market maker).  
    These events are modelled 
    as an independent Poisson process with rate $\mu$ 
    (number of LT arrivals per time unit).  
    \end{enumerate}

   \item {\bf Detailed chronology of the MZ game} \quad  
The description below is based on the MZ paper, but 
attempts to clarify the process in more detail 
(cf. Fig~\ref{fig:mz_game_chronology}). 

\begin{figure}[htbp]
    \centering
     \includegraphics[scale=0.45]{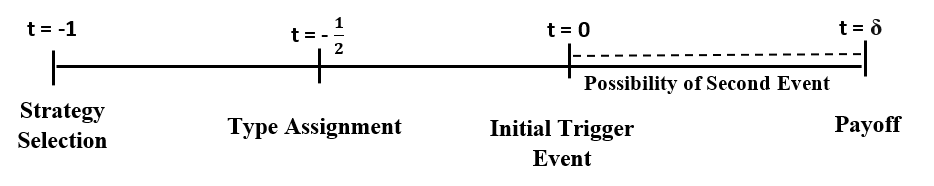}
     \caption[caption]{Chronology of the MZ game}
    \label{fig:mz_game_chronology}
\end{figure}

\begin{itemize}
    \item \underline{At $t=-1$,} {\bf Strategy 
    selection:} \quad all HFTs are informed 
    about the game's parameters (i.e. $\alpha, \mu, \delta, \sigma$ 
    and $H$).  Based on this information they can then compute  the 
    expected utility as a function 
    of the spread $s$ (and the sniping behaviour) for both the market maker and bandits 
    (more details in section~\ref{sct:pay-off}), but are still 
unsure about which one applies to them. Taking this uncertainty 
into account, 
each HFT 
decides on a spread $s$, and 
posts a single bid and ask quote in an empty order book 
(i.e.  initially each HFT has a zero position\footnote{A zero position is when traders don't own any amount of financial assets or commodity}).

 \item \underline{At $t=-1/2$,}  {\bf Type assignation:  } \quad one of the HFTs is chosen to be the market maker, 
whereas all the others will become "bandits" 
(eager to snipe). 
The selection process of the market maker proceeds in two steps: 

\begin{enumerate}
    \item First,  the HFT(s) that posted the smallest spread $s$ are 
    selected;
   \item If there are multiple HFTs that posted the same minimal 
   spread, then one of them is selected at random (uniformly). 
\end{enumerate}
The fact that the outcome of this 
selection process is uncertain, 
is an important aspect of the game that the HFTs need to take 
into account when selecting a strategy at $t=-1$ as it is 
impossible for them to change their position at 
a later time.

\item \underline{At $t=0$, } {\bf Initial trigger 
starts game: }\quad 
one of two possible {\it trigger events} occurs: 

\begin{enumerate}
 \item {\it News event} (public) that changes the value of the asset 
 (i.e. value shock of $\pm \sigma$). This immediately triggers 
 a race among the HFTs: the market maker will try to update 
 his stale quotes, while the bandits will attempt to snipe. 
 The winner of this race is randomly (uniform) chosen 
 among all contestants. 
 The winner is known at time $t= \delta$.
   
\item {\it LT arrives at matcher (queue)}. This will result 
    in a transaction, which is invariably profitable for the market  
    maker as he will cash in the spread ($s$).  This event does not 
    elicit any reaction from the HFBs. 
\end{enumerate}

\item \underline{Interval $0 < t < \delta$} \quad  During this 
interval one (and only one) of three things can happen: 

\begin{enumerate}
    \item An additional news event becomes (publicly) known 
    (rate $\alpha$). 
    \item Another LT arrives at the matcher, 
    intent on interacting with the market maker (rate $\mu$);
    \item Nothing happens. 
\end{enumerate}

\item \underline{At $t=\delta$,} {\bf Conclusion and pay-off: } 
\quad The game concludes 
and the positions are used to compute the pay-off 
and 
corresponding utility for 
both the bandits and market maker (see section~\ref{sct:pay-off} 
for more details).  
Importantly, the utility of  a negative 
pay-off is further inflated by the {\bf risk-aversion factor} $\gamma$:
\begin{equation}
\mbox{utility }  = 
\left\{   
\begin{array}{rcl}
    \makebox{pay-off}& \quad& \mbox{if pay-off $\geq 0$}  \\
    \gamma \cdot \makebox{pay-off}& & \mbox{if pay-off $< 0$} 
\end{array}
\right. 
\label{eq:neg_payoff_inflation}
\end{equation}
where $\gamma \geq 1$. 

\end{itemize}

\end{itemize}
In the next sections we will explore some of 
the concepts in more detail, starting with the detailed 
computation of the pay-off.

\subsection{Payoff and utility computation}
\label{sct:pay-off}
\subsubsection{General principles}
\label{sct:payoff_computation}
To compute the payoff for specific game-outcomes 
for each HFT we observe that there are three sources of 
payoff: 

\begin{itemize}
   \item {\it Changes in position}: \quad 
   obtaining more (or less) of a financial asset results in 
   a corresponding change in wealth; 
   \item {\it Change of the intrinsic value of an asset: } if the intrinsic 
   value of an asset changes (due to good or bad news), the owner benefits (or suffers)
   accordingly; 
    \item {\it Income} \quad from a successful transaction 
    (either buy or sell). 
  The income can be both positive or negative depending on 
    whether the HFT gets paid or needs to pay to complete the   transaction.
\end{itemize}
These observations can be summarized in the following equations: 
 \begin{align*}
   \mathbf{payoff} &=   \underbrace{\mathbf{position}_{(t=\delta)} \times \mathbf{value}_{(t=\delta)}}_{\mbox{final}} \,\, + \,\, \underbrace{\mathbf{income}}_{\mbox{during}} - \underbrace{\mathbf{position }_{(t=0)} \times \mathbf{value}_{(t=0)}}_{\mbox{initial}} \\
    &
\end{align*}
but since we are assuming that all HFTs start with a zero position, the last term vanishes 
 and we therefore get the simplified formula: 
  \begin{align}
    \mathbf{payoff}  &=  \mathbf{position}_{(t=\delta)} \times \mathbf{value }_{(t=\delta)} \,\, + \,\, \mathbf{income}_{(t=\delta)} &
    \label{eq:payoff_computation_1}
\end{align}
The corresponding utilities $ U_M$ (for the market maker) 
and $U_B$ (for a bandit) can then by inflating 
negative payoffs with the risk aversion factor $\gamma$ (cf. 
eq.~\ref{eq:neg_payoff_inflation}).

In addition to the primary MZ-parameters 
introduced above 
(viz. $H, \alpha, \mu, \delta$ 
and $\gamma$) we will introduce some 
useful 
shorthand notation for 
derived quantities that recur throughout 
this paper: 

\begin{itemize}
    \item $\overline{\alpha} := \alpha \delta/2$ and 
    $\overline{\mu} := \mu\delta/2$ are (half) the expected number 
    of news and  LT arrivals, respectively, in the interval between 
    the start of game ($t=0$) and its conclusion (at $t=\delta$). 
    Since we are assuming that $\delta$ is sufficiency 
    small so that we expect less than one event to happen 
    in each interval, we have the 
    following strict inequality:
    
    \begin{equation}
        (\alpha + \mu) \delta < 1  \quad \quad \mbox{or again} 
        \quad \quad \overline{\alpha} + \overline{\mu} <  \frac{1}{2}.
        \label{eq:cond_expectation_2nd_event}
    \end{equation}
       
     \item $\beta := \alpha/(\alpha+\mu)$ is the probability 
        that the trigger event  will be arrival of news 
        rather than an LT arrival.  Obviously, the 
        latter therefore has probability $1-\beta$. 
    \item Some additional notation to streamline the equations 
    below: 
    \begin{equation}
     q:=\gamma-1 \quad\mbox{and}\quad  m = 1-\overline{\mu}
    \quad\mbox{and}\quad \overline{\theta} 
    = 2\,\frac{\overline{\alpha}\,\overline{\mu}}{\overline{\alpha} + \overline{\mu}}.
    \label{eq:q_harmonic_mean}
    \end{equation}
    Notice that  $\overline{\theta}$ is the harmonic mean of 
     $\overline{\alpha}$ and  $\overline{\mu}$ 
     therefore tends to be closer to  
     the smaller of the two.

\end{itemize}

\subsubsection{Payoffs for specific event-sequences} 
\label{sec:payoff_specific_sequence}
Since the chronology of the MZ game is strictly defined, 
there 
are only 20 possible event sequences that can 
happen between the initial trigger 
(at $t=0$) and the final conclusion 
of the game at $t=\delta$.  
All the possible combinations and their corresponding payoffs (and utilities) are 
given in  \cite{mz2017} and 
summarized for the reader's convenience 
in Table~\ref{tab:Pay-off_referring_to_table1.p1201}. It is important to realise that bandits 
can only benefit from the game if their sniping 
attempt is successful. Hence, if there is no race (i.e. trigger 
event is the arrival of a liquidity trader), or if a bandit  loses the 
race, then their payoff equals zero. Furthermore, 
if a race ensues which the market maker loses, {\it only 
the successful sniper} will gain utility 
(2nd utility column in table). 

For a detailed explanation of each event's payoff and probability 
we refer to the appendix \ref{sct:payoff_details}, but we will illustrate the general ideas 
using {\tt NG-LA} (first event is good news and the second event is  arrival of LT on ask) as a specific example:

\begin{itemize}
    \item {\bf Probability of event:} \quad Since,  
    in general, the first (trigger) event is either 
    the publication of a news item (rate $\alpha$) or  the arrival of 
    a liquidity trader (LT, rate $\mu$), the probability of the former 
    equals $\beta := \alpha/(\alpha + \mu)$. Since for half of these events 
    the news is good, the probability of the trigger event being {\tt NG} 
    equals $\beta/2$. The probability of a second news event arriving 
    in the interval $[0, \, \delta) $  equals $\alpha \delta$ and if we 
    insist on this news item also being good, the probability halves to 
    $\alpha \delta/2 = \overline{\alpha}$. Since the two events are independent, 
    we can simply multiply the two probabilities. 
    
    \item {\bf Payoff of event:} \quad 
    In the case of NG-LA, the first (i.e. trigger) event is good news and hence increases the intrinsic value of the asset (by an 
    amount $\sigma$). As a 
    consequence, the market maker (MM) will race 
    to the exchange to cancel his (now stale) 
    order in order to prevent loses, 
    while all other HFTs (bandits)  
    race to snap up the higher value asset 
    at the outdated low price (sniping).

    The second  event is the arrival of LT on {\it ask}  (value +s) and means 
    that an uninformed liquidity trader (LT) 
    happens to join the queue 
    at the matching engine 
    ahead 
    of all the HFTs (who are still racing) and 
    manages to buy the asset at the outdated 
    (low) ask price.  This is of course bad 
    news for the market maker who ends up  
    selling an asset below its current 
    market value. The payoff for all the bandits 
    is zero as they failed to snipe. 
    By referring to equation~\eqref{eq:payoff_computation_1}, we can compute payoff for market maker:

\begin{eqnarray}
    \mathbf{payoff}_{M}  &= & \mathbf{(position)}_{(t=\delta)} \times \mathbf{(value)}_{(t=\delta)} \,\, + \,\, \mathbf{income} \nonumber\\
    &=& (-1) \times (v+\sigma) \quad + \quad(v+s)  \nonumber \\
    &= & - (\sigma - s)
\end{eqnarray}
Since the payoff is negative, the 
corresponding utility ($u_M$) is inflated by the 
risk aversion factor $\gamma$, hence:

\begin{equation}
u_M = -\gamma (\sigma - s)   \quad\quad \mbox{and} \quad\quad
u_B = 0.
\label{eq:util_NG_LA}
\end{equation}

\end{itemize}

 \begin{verbatim}
For more details, see:
KPD_supplementary_material_1.ipynb (section.1)
\end{verbatim}

\begin{table}[htbp]
    \centering  
  \begin{tabular}{|l|l|l|c|c|c|c|}
  \hline
     & &\multicolumn{5}{c|}{{\bf Utilities}}   \\  
  \hline
   & & &\multicolumn{2}{c|}{MM loses race } &\multicolumn{2}{c|}{MM wins race }  \\  
    & & &\multicolumn{2}{c|}{(Prob = $h$)} &\multicolumn{2}{c|}{ (Prob = $1-h$)}    \\   \hline
   \hline
{\bf     Event code} & \multicolumn{2}{c|}{1st and 2nd event prob}& MM & B (sniper!) & MM & B (all!)  \\ \hline
    NG-NG & $\frac{1}{2}\beta$ & $ \overline{\alpha}$ & $  -\gamma(2\sigma-s)$  &    $ 2\sigma-s $ &     $0 $  &     $0 $    \\\hline
    NG-NB  &  $\frac{1}{2}\beta $ &  $\overline{\alpha}$  &  $s$         &    $-\gamma s$  &  $0 $ &   $0 $\\\hline
    NG-LA & $\frac{1}{2}\beta $ &  $\overline{\mu}$ & $-\gamma(\sigma-s)$      &      $0 $   &$-\gamma(\sigma-s)$&   $0 $  \\\hline
    NG-LB   & $\frac{1}{2}\beta $ & $ \overline{\mu}$ &   $2s  $     &    $\sigma-s$ &  $\sigma+s$   &   $0 $  \\\hline
       NG-no  &  $\frac{1}{2}\beta $ &  $1- 2(\overline{\alpha} +\overline{\mu})$  &    $-\gamma(\sigma-s)$      &   $\sigma-s$   &    $0 $   &  $0 $  \\ \hline
   NB-NG & $\frac{1}{2}\beta $ & $ \overline{\alpha}$ &  $s$         &    $-\gamma s$  &  $0 $ & $0 $ \\\hline
   NB-NB & $\frac{1}{2}\beta $ &  $\overline{\alpha}$&  $-\gamma (2\sigma -s)$ &   $2\sigma-s$ &   $0 $ &   $0 $ \\\hline

 NB-LA&   $\frac{1}{2}\beta $ &  $ \overline{\mu}$ &   $2s$      &    $\sigma-s$ &  $\sigma+s$   &    $0 $     \\\hline

 NB-LB &  $\frac{1}{2}\beta  $ &  $ \overline{\mu}$ &   $-\gamma (\sigma-s)$      &     $0 $ &  $-\gamma (\sigma-s)$   &     $0 $  \\\hline
    
NB-no & $\frac{1}{2}\beta $ & $ 1- 2(\overline{\alpha} +\overline{\mu})$ &      $-\gamma(\sigma-s)$      &   $\sigma-s$  &    $0 $   &   $0 $ \\ \hline 
\hline
 & & & \multicolumn{2}{c|}{No race!} &  \multicolumn{2}{c|}{}\\ \hline
 & & & MM & B (all!) & -&- \\ \hline
     LA-NG &$\frac{1}{2}(1-\beta) $ & $ \overline{\alpha}$ &  $-\gamma(\sigma-s)$      &      $0 $   & - & -     \\\hline
    LA-NB  &$\frac{1}{2}(1-\beta) $&   $\overline{\alpha}$  &    $\sigma+s$     &       $0 $ &- & -  \\\hline
    LA-LA   &  $\frac{1}{2}(1-\beta) $&  $ \overline{\mu}$         &  $s$           &         $0 $    &  -   &  - \\\hline
    LA-LB   & $\frac{1}{2}(1-\beta) $ &  $\overline{\mu}$        &$2s$     &        $0 $   &  -   & - \\\hline
    LA-no  &$\frac{1}{2}(1-\beta)  $&  $ 1- 2(\overline{\alpha} +\overline{\mu})$ & $s$      &      $0 $   &   - &-  \\\hline
    LB-NG    &  $\frac{1}{2}(1-\beta) $ & $ \overline{\alpha}$       &  $\sigma +s$     &   $0 $  & - &-  \\\hline
    LB-NB    &  $\frac{1}{2}(1-\beta) $ &  $\overline{\alpha}$      & $-\gamma(\sigma-s)$      &  $0 $   &     -    &   -  \\\hline
    LB-LA  &$\frac{1}{2}(1-\beta) $ &   $\overline{\mu}$ &  $2s$ &       $0 $ & - & -  \\\hline
    LB-LB     & $\frac{1}{2}(1-\beta)  $ &  $\overline{\mu}$      &   $s$          &    $0 $ & -  &  -   \\\hline
       LB-no   &  $\frac{1}{2}(1-\beta) $ &  $1- 2(\overline{\alpha} +\overline{\mu})$        & $s$          &     $0 $&  -  &  -  \\\hline
       \hline 
        {\bf  Legend}&       \multicolumn{6}{c|}{NG/NB: Good/Bad News, LA/LB = liquidity trader on ask/bid,  no = no event}\\
 \hline
\end{tabular}
\caption{Payoff table for all possible events in MZ-game. 
For instance, in NG-LA, the first event is good news and the second event is  arrival of LT on ask.
The negative payoffs are made explicit using a minus sign 
and are inflated with the risk aversion factor $\gamma \geq 1$
in order to obtain the actual utility.  
The trigger event for the first ten cases is the publication of a news 
item which sets off a race. A bandit can only receive payoff if he 
succeeds in winning that race, in all other cases there is no payoff. 
For that reason we only list the utility of the bandit that was successful 
as a sniper. For the explanation of the probabilities that the 
market maker will lose or win the race, we refer the reader to 
section \ref{sct: utility_computation}). 
Notice that for the last ten events 
for which the trigger event is the arrival of liquidity trader, there is 
no race, and so no need to distinguish between the 
possible fates of the market maker. 
For detailed computation of the table, see appendix \ref{sct:payoff_details}.}
    \label{tab:Pay-off_referring_to_table1.p1201}
\end{table}

\clearpage

\subsubsection{Expected utilities} 

Table~\ref{tab:Pay-off_referring_to_table1.p1201} lists 
the utilities of specific events 
for both the market maker and any of the 
bandits once we specify  the spread $s$.  
However, in the actual game these events are generated 
according to a stochastic process 
and the utilities are therefore stochastic 
variables (denoted by capital letters $U_M$ and 
$U_B$). To compute the 
{\it expected} utilities ($\E U_M$ and $\E U_B$)  
we need to sum the  utilities 
of all relevant events  weighted by the corresponding probabilities.  
The event probabilities are listed explicitly in 
the table. 
But we also have to take into account  the probabilities with which any of the HFTs  
will win or lose the race 
as this changes their pay-off. 
Below we explore this in more detail 
for the market maker and bandit separately. 

\paragraph{Market maker} 

A glance at 
Table~\ref{tab:Pay-off_referring_to_table1.p1201} shows that the market maker 
receives non-trivial utility for all  the 
events.  
To streamline the exposition  
we will introduce a wildcard and denote  all event codes that correspond 
to a  news trigger event as {\tt N***}.  
Each of these events 
occurs with probability $\beta/2$ and provokes 
a race.  Similarly, events triggered by 
an LT arrival will be denoted as {\tt L***} 
and occur with probability $(1-\beta)/2$ but 
don't spark a race. 
Furthermore, when there is an actual race, 
the market maker's utility depends 
on whether or not he beats the other HFTs to the 
finish.  To take this into account,  
we introduce $h$ to denote the probability 
that the market maker will {\it lose} the race.  
  Since all the HFTs 
enter the race and have 
the same probability of winning it, 
we conclude: 
\begin{equation}
h := P\left\{\mbox{MM {\bf loses} race }\given 
\mbox{there is a race}\right\} = 
\frac{H-1}{H}.
\label{eq:h_orig}
\end{equation}
Gathering all this information 
we can now express the expected utility 
of the market maker as a weighted sum 
over the possible {\it second } event $e_2$ 
where $p(e_2)$  and $u_M(e_2)$ are the corresponding {\it event}
probability and utility, respectively: 

\begin{eqnarray}
 \E U_M(s) & =& \frac{\beta}{2} 
\left[ \sum_{e_2 \in N***} p(e_2)\,\left\{h \, u_M(e2\given \mbox{MM loses}) 
+ (1-h) \, u_M(e_2\given \mbox{MM wins}) \right\}\right] \nonumber\\
&& + \, 
\left(\frac{1-\beta}{2}\right) \left[\sum_{e_2 \in L***} p(e_2) \,u_M(e_2) \right]. 
\label{eq:EU_M_schematically_1}
\end{eqnarray}
%
%
In section \ref{sct: utility_computation} we will see how we need to re-interpret this 
expression in the case of probabilistic sniping.

\paragraph{Individual bandit}   The expected 
utility for an {\it individual} bandit can be computed 
along similar lines.  In fact, 
the calculation is easier as a 
bandit only receives payoff if his sniping attempt is 
successful. 
If there is a race (event code {\tt N***}), the 
probability that a specific, individual bandit 
will win equals 
\begin{equation}
    g :=P\left\{\mbox{bandit {\bf wins} race }\given 
\mbox{there is a race}\right\} = \frac{1}{H}
\label{eq:g_orig}
\end{equation}
as there are $H$  agents in the race.  
Notice that this is identical to the probability $1-h$ that 
the market maker will win the race -- for the obvious 
reason.  
It might therefore seem an unnecessary complication 
of the notation, but  once we introduce the 
concept of probabilistic sniping, these two 
quantities will diverge in meaning. 

Since all {\tt L***} events result in zero utility 
for a bandit (no race),  we only need to sum 
over the {\tt N***}  events, resulting in the 
following expression:

\begin{equation}
    \E U_B(s) = \frac{\beta}{2}  g \sum_{e_2 \in N***} 
    p(e_2) \, u_B(e_2 \given \mbox{MM loses race})
\quad \quad  \mbox{where } \quad g = \frac{1}{H}. 
\label{eq:EU_B_schematically_1}
\end{equation}
\begin{verbatim}
See KPD_supplementary_material_1.ipynb (section.2.1)
\end{verbatim}

\paragraph{Some simplifications}

Expanding eqs.~\ref{eq:EU_M_schematically_1} and 
\ref{eq:EU_B_schematically_1}
is 
straightforward but tedious and 
therefore deferred to appendix \ref{appx:expansion_expected_util}.
However,  even without plowing through the algebra, 
a cursory inspection of 
table \ref{tab:Pay-off_referring_to_table1.p1201} shows 
that  all event utilities  
    are proportional to either the spread $s$ or 
the difference ($w\sigma \pm s, \, \mbox{where } w = 1,2$) 
of spread and jumpsize $\sigma$.  From this, we can 
introduce the following notational simplifications:

\begin{itemize}
    \item {\bf Rescaling} \quad  We can factor out $\sigma$ by dividing both 
the spread $s$ and  utility $U$  by $\sigma$.  This 
amounts to rescaling both spread and utility by using 
the jumpsize $\sigma$ as a natural yardstick: 
$$ s \longrightarrow \tilde{s}:= \frac{s}{\sigma}  \quad\quad
\mbox{and} \quad\quad U \longrightarrow 
\tilde{U} := \frac{U}{\sigma}$$
This rescaling allows us to fix $\sigma \equiv 1$ which slightly simplifies 
the equations in the remainder of the paper without having any 
impact on the qualitative nature of the conclusions. 
Notice that we will 
drop the tildes for notational convenience, and simply 
use $s$ and $U$ to denote spread and utility measured 
using $\sigma$ as unit. 

\item {\bf Utilities are linear in $s$}  \quad  The expected 
utilities for both the market maker ($\E U_M$) and bandit ($\E U_B$) 
are linear combinations of the utilities in the table and therefore 
result in linear (or more precisely, affine) functions in the 
spread $s$: 
\begin{equation}
 \E U_B(s) =  A(1-s)  + Bs, \quad \mbox{where }  A := \E U_B(s=0) 
\quad \mbox{and } B :=  \E U_B(s=1),  
\label{eq:AB_intro}
\end{equation}
and similarly: 
\begin{equation}
 \E U_M(s) =  C(1-s)  + Ds,  \quad \mbox{where } \quad C := \E U_M(s=0) 
\quad \mbox{and } D :=  \E U_M(s=1) .
\label{eq:CD_intro}
\end{equation}
Hence, the values $A, B, C$ and $D$ are the endpoints of 
the linear line-segments that represent the expected utilities 
for the market maker and a bandit as a function of spread $0 \leq s \leq 1$ 
(also see Fig.~\ref{fig:intersection_utilities}). 
Expressing the utilities in terms of these endpoints is useful 
as it simplifies some the formulae in the remainder of this paper. 

\end{itemize}

\subsection{Computing the point of 
indifference}
\label{sct:computing_equilibrium}

\begin{figure}[htbp]
    \centering
    \includegraphics[width=0.8\textwidth]{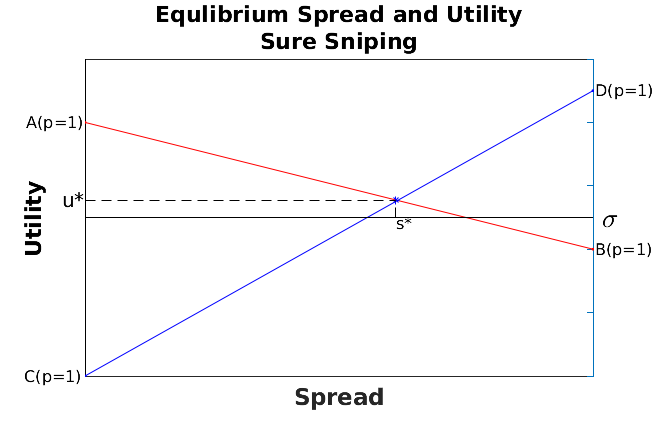}
    \caption{Graph representing the expected utilities 
    $\E U_B(s)$ (line AB) 
    and $\E U_M(s)$ (line CD) for a bandit and the market maker,  respectively. 
    The HFTs will select spread $s^*$ (resulting in expected 
    utility  $u^*$) 
    by determining the intersection of the 
    two utilities (point of indifference).  
    For the time being, we only consider sure sniping which corresponds 
    to $p=1$ .  For more details on probabilistic sniping 
    (for which $p<1$), see section~\ref{sct:prob_vs_sure}. 
    }
    \label{fig:intersection_utilities}
\end{figure}

Recall from the description of the MZ-game in 
section~\ref{sct:mz-game_ingredients} that 
at the inception of the game (i.e. at $t=-1$ when 
they need to specify the spread $s$ for their quotes), 
the HFTs are agnostic 
about their eventual role in the game as this is only  
assigned at $t=- \frac{1}{2}$.  
The intersection point  between 
the two utilities will therefore 
play a pivotal role in the HFT's decision making as it represents the spread 
(hereafter denoted by
$s^*$)  that makes them indifferent as 
to which role they will play. 
In terms of the quantities $A,B, C$ and $D$ as 
defined above, it is  straightforward to 
determine  $s^*$ and the 
corresponding utility $u^* := \E U_B(s^*) \equiv \E U_M(s^*)$ 
by computing the intersection of the 
linear utility functions (also see 
Fig.~\ref{fig:intersection_utilities}): 
\begin{equation}
  s^* = \frac{(A-C)}{(A-C)+(D-B)}  
  \quad\quad \mbox{and} \quad\quad
   u^* = \frac{(AD-BC)}{(A-C)+(D-B)}
    \label{eq:u_star_sure}    
\end{equation}
It is shown in \cite{mz2017} that,  if 
$u^*>0$ (intersection lies above x-axis), 
the MZ game has a unique {\bf pure Nash equilibrium } 
in which all HFTs will pick the optimal spread $s^*$ and 
expect to earn the utility $ u^*$. 
\begin{verbatim}
For more details,see: 
KPD_supplementary_material_numeric.ipynb (section.2)
\end{verbatim}

However, if $u^* \leq 0$ (intersection lies on or below x-axis), things are less straightforward as pursuing the above strategy will now result in losses for 
both the market maker and the bandits. 
One of the contribution in \cite{mz2017} is the realisation that probabilistic 
sniping might suggest an additional equilibrium. 
This will be explored in detail in the next section.

\section{Probabilistic versus Sure Sniping}
\label{sct:prob_vs_sure}
\subsection{The rationale for probabilistic sniping} 
\label{sct:rational_prob_sniping}
Up till now we have  assumed that all bandits will race to the exchange 
whenever an opportunity arises (i.e. whenever  
the trigger event is the publication of a  news item). 
This seems reasonable as a successful snipe is the only way 
a bandit can gain positive utility.  However, a careful inspection 
of 
Table~\ref{tab:Pay-off_referring_to_table1.p1201} reveals that 
sniping has a potential down-side. For instance, 
if a positive news fact 
is followed by a negative one (NG-NB), or vice versa (NB-NG), 
the sniper acquires an 
asset that has no added value and, as a consequence, incurs negative 
utility. Hence, under these circumstances it might be more profitable 
to refrain from sniping, or to snipe only occasionally. This is 
translated into the idea of {\bf probabilistic sniping} in which 
an individual bandit enters the race with 
{\bf sniping probability} $p$. 

\begin{figure}[htbp]
\centering
   \includegraphics[width=0.7\textwidth]{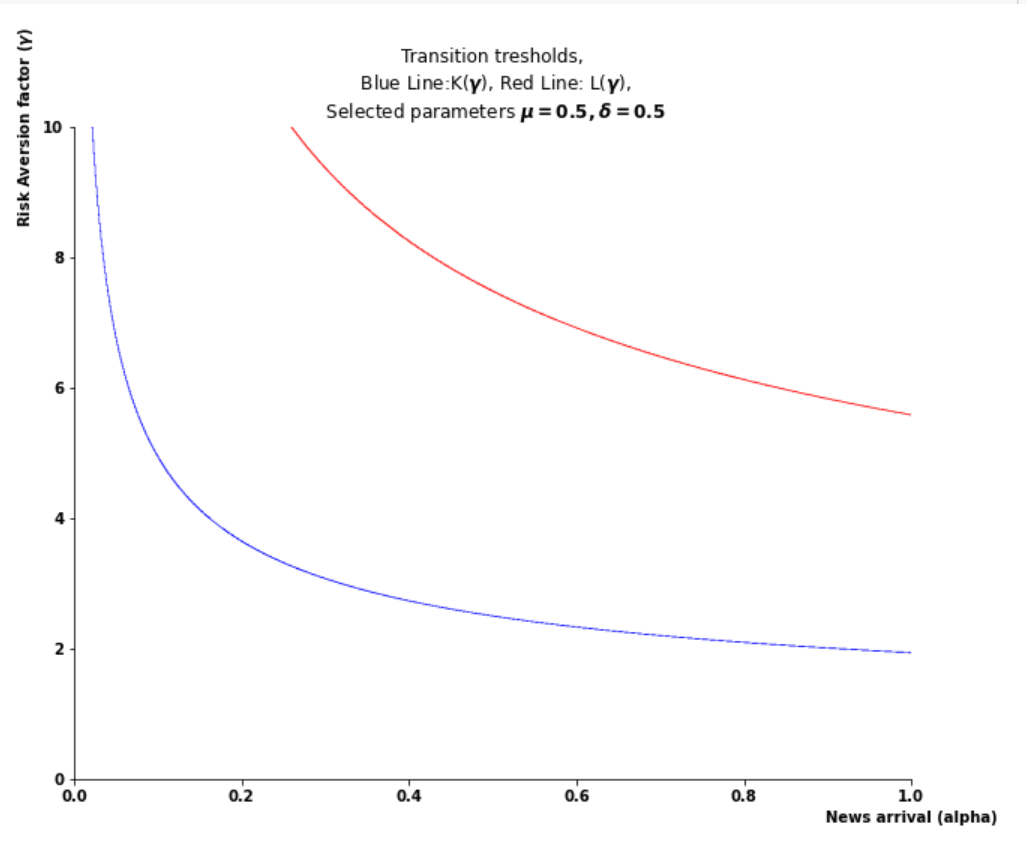}\\
\includegraphics[width=0.7\textwidth]{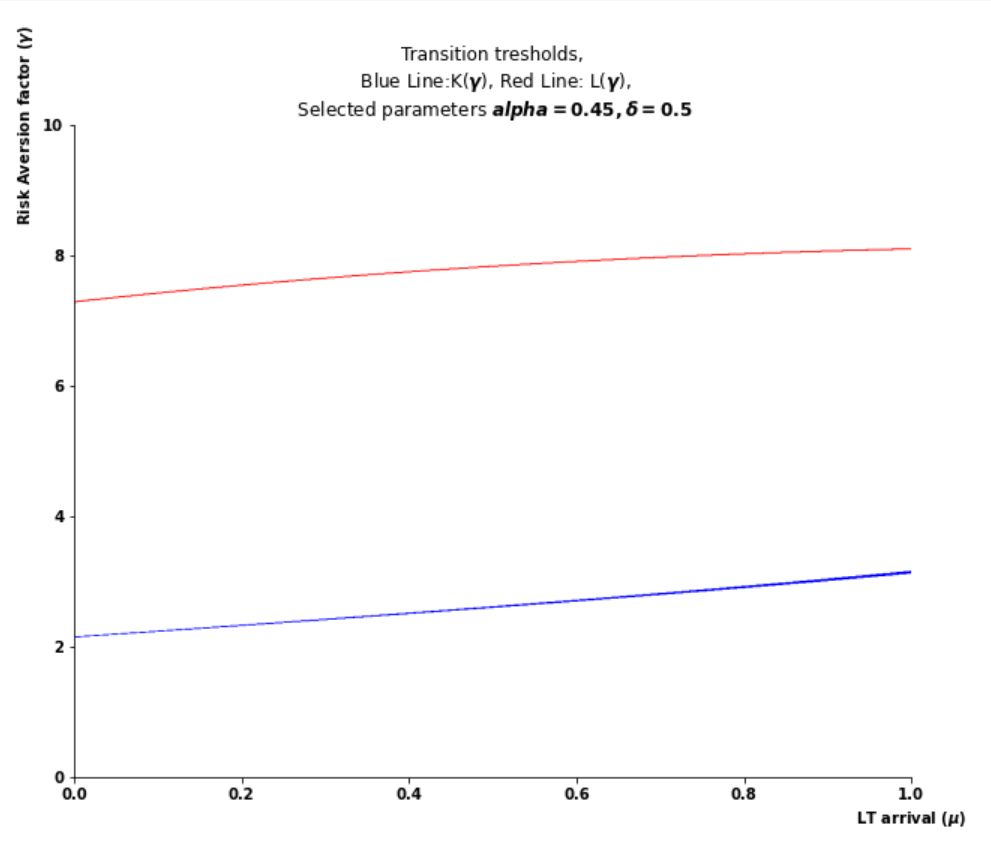}
\caption{These figures show the evolution  of transition thresholds $\overline\gamma_K$ (blue) 
and $\overline\gamma_L$ (red) as functions of  
the rates for the trigger 
event: arrival rate  of news ($\alpha$, TOP) and arrival rate of LT ($\mu$, bottom).  
Notice (top) how increasing the news rate $\alpha$ results a faster 
transition to probabilistic sniping (as explained in 
section~\ref{sct:rational_prob_sniping} ). 
 }
  \label{fig:threshold_dependency_alpha_mu}
\end{figure}

Obviously,  less sniping is always  advantageous 
for the market maker as 
sniping reduces his payoff. 
So it seems that there are situations (e.g. high news rates $\alpha$, 
see caption of 
Fig.~\ref{fig:threshold_dependency_alpha_mu})
for which 
probabilistic sniping could actually be advantageous for all parties. 
In the next sections we will quantify in more detail how  
probabilistic sniping affects the various utilities.

\subsection{The impact of probabilistic sniping } 
\label{sct: utility_computation}

We start by defining more 
formally the sniping probability $p$ to be 
    the probability that an individual bandit will participate 
    in the race, given that there is a race,  
    i.e. given that the initial trigger event is "news":
    \begin{equation}
    p:= P\left\{\mbox{Individual bandit will race} \given 
    \mbox{Tr(igger) = N(ews)}\right\}
        \label{def:prob_sniping}
    \end{equation}
Obviously, $p=1$ in the case of sure sniping.  
Notice also that 
this probability is only relevant for bandits, as the market 
maker will always enter the race! 
The corresponding change in sniping behaviour implies that we have to 
re-evaluate the probabilities $h$ (probability that the market 
maker will lose the race) and $g$ (probability that 
an individual bandit who entered the race, will 
actually win it) 
as both quantities will now depend on $p$.  

\begin{enumerate}
    \item {\bf Probability $h(p)$ that market maker will lose the race}  \quad 
In the case of sure sniping, all HFTs enter the 
race and since they all have equal probability of winning, the probability that the market maker will {\it lose} equals 
$h =(H-1)/H$.  
However, when individual bandits snipe (independently!) with a probability $p<1$, 
 the probability that the market maker will get sniped 
 depends on the number of bandits that do attempt to out-race him, 
 and therefore  depends on $p$.  We will therefore 
denote it as $h(p)$, defined  formally  as: 

\begin{equation}
 h(p):= P\left\{\mbox{MM will lose MZ race} \given 
 \mbox{individual bandits enter race with prob $p$} \right\}
\label{eq:def_hp}    
\end{equation}
A straightforward calculation (see Appendix \ref{sct:computation_h_and_g}) shows that 
%
\begin{equation}
    h(p) = \frac{(1-p)^H - (1-Hp)}{Hp}
    \label{eq:hp_compact}
\end{equation}
which indeed reduces to $h \equiv h(1) = (H-1)/H$ in the case of sure 
sniping ($p=1$).  Furthermore, using the 
expansion
\begin{equation}
\
(1-p)^H = 1 -Hp + \frac{H(H-1)}{2} p^2 - \frac{H(H-1)(H-2)}{3!} p^3 + \ldots +(-p)^H  
\label{eq:h_asymptotics_1}
\end{equation} 
shows that we have the following asymptoptics 
for small values of $p \downarrow 0$: 
\begin{equation}
h(p) = \left(\frac{H-1}{2}\right) p + o(p^2),  
\label{eq:h_asymptotics_2}
\end{equation}
whence $\lim_{p\downarrow 0} h(p) = 0$. 
The probability $h(p)$ plays a crucial role in the utility computation 
for both the market maker and the bandits. 

\item {\bf Conditional probability $g(p)$ that an individual bandit wins race, given he enters race}  
In the case of probabilistic sniping, the probability that 
an individual bandit will win the race needs to be 
conditioned on him actually entering the race, as this is 
no longer a sure thing. Therefore, we introduce the 
following formal definition of this conditional probability:

\begin{equation}
g(p) := P\{ \mbox{individual bandit will  {\it win} race} \given  
 \mbox{he participates in the race} \}
    \label{def:g(p)}
\end{equation}
The actual computation  $g(p)$ is deferred to 
Appendix \ref{sct:computation_h_and_g}   
but yields:  
%
\begin{equation}
g(p) = \frac{(1-p)^H - (1 -pH)}{H(H-1)p^2} =  \frac{h(p)}{(H-1)p}.
    \label{eq:g(p)}
\end{equation}
Intuitively, this result makes sense: $h(p)$ is the probability 
that a bandit rather than the market maker will win the race, 
which is then divided by $(H-1)p$, 
i.e. the expected number of bandits in the 
race. 

Another way to see the connection between 
$h(p)$ and $g(p)$ is to observe that $pg(p)$ 
is the {\bf un}-conditional probability that 
an individual bandit will win the race. For this 
to happen, the market maker needs to lose the 
race (probability $h(p)$)  and each of the $ (H-1)$ bandits has the 
same odds.  Hence: 
\begin{equation}
    p g(p) = \frac{h(p)}{H-1}
    \label{eq:h_and_g_alternative}
\end{equation}

Notice also that if every bandit races for sure ($p=1$), then the  probability that 
an individual bandit will win, simplifies to  
$g \equiv g(1) = 1/H$ as expected. 
In the same vein, using the asymptotics in 
eq.~(\ref{eq:h_asymptotics_2}), 
it follows that 
\begin{equation}
    \lim_{p\downarrow 0} g(p) = 1/2.  
    \label{eq:g_asymptotics}
\end{equation}
This also makes sense: if $p \approx 0$ then any bandit that is in the race will 
probably be the only bandit in that race, and therefore 
will have a probability of 1/2 to outrun the racing market maker.

\end{enumerate}
%

 \begin{verbatim}
For more details, see:
(KPD_supplementary_material_2.ipynb)
\end{verbatim}

\subsection{Expected utilities under probabilistic sniping}
\label{sct:util_prob_sniping}

Now that we have a precise definition of the probabilities 
$h(p)$ and $g(p)$ we are in a position to  determine 
the expected utilities for both the market maker and  
individual bandits under probabilistic sniping. 

\subsubsection{Expected utility for bandit}

As pointed out above, the bandit will only race if the trigger event ($Tr$)
is the arrival of a news item ($N$). 
Recall that this happens with probability $\beta = \alpha/(\alpha+\mu)$. Hence, we can compute the 
expected utility by conditioning on the trigger event being news.  
Furthermore, we need to take into account that the bandit will 
only participate in the race with probability $p$, and needs 
to win the race in order to gain utility.  We therefore find: 

\begin{eqnarray*}
     \E U_B &=& \E(U_B \given Tr = N)\, P(Tr = N) \nonumber \\
     &=&   \E(U_B \given Tr = N) \,\beta\\
     &=&   \E(U_B \given Tr = N \, \& \, \mbox{bandit enters race}) 
     \, P\{\mbox{bandit enters race}\}\,\beta\\
     &=&  \E(U_B \given Tr=N \,\&\,\mbox{bandit enters and wins race}) 
     P\{\mbox{bandit wins race} \given \mbox{he races} \} \, p \beta\nonumber\\
     & =& \E(U_B \given Tr=N \,\&\, \mbox{bandit enters and wins race}) \, g(p) p \beta
      \end{eqnarray*}
From this we conclude:
\begin{equation}
 \E U_B(s,p) =  \beta g(p) p\, \E (U_B \given Tr=N\,\&\, \mbox{bandit enters and wins race})
    \label{eq:U_B_0}
\end{equation}
To further elaborate the expression for $\E U_B$  we need to condition the RHS on 
the second  event.  
This amounts to summing the utilities (weighted with the probabilities 
of the second event) in the sniper column of Table, which  
yields:

and a straightforward but tedious 
calculation shows:  

$$
\E (U_B \given \mbox{Tr = N \quad \& \quad bandit enters and wins race})  = (1-\overline{\mu})(1 - s)- 
\overline{\alpha} q s. 
$$
Plugging this into eq.~\ref{eq:U_B_0} shows how 
the expected utility of each bandit depends 
on spread $s$, sniping probability $p$ and 
risk aversion $\gamma \geq 1$:

\begin{equation}
    \E U_B(s,p, \gamma) =  \beta p g(p) 
    \left\{(1-\overline{\mu})
(1 - s)- \overline{\alpha} (\gamma-1) s \right\} 
=  \beta p g(p) 
    \left\{ m (1 - s)- \overline{\alpha} q s \right\}
.
\label{eq:EU_B_1}
\end{equation}
In particular, this means that the end points of the line segment, as  
functions of $p$ and $\gamma$, are given by: 
\begin{equation}
\begin{array}{lclclcl}
 A & \equiv & A(p) &:=& \E U_B(s=0) &=& m \beta p g(p) \\
 B &\equiv & B(p,\gamma) &:=& \E U_B(s =1)  &=& -\overline{\alpha}(\gamma-1) \beta p g(p)
 \end{array}
 \label{eq:A_and_B_fion_of_p_and_gamma}
 \end{equation}

We conclude this paragraph with a 
straightforward but useful  observation. 
The value $s^{**}$ for which the bandit's utility 
becomes zero (i.e. the intersection point with the x-axis, see Fig.~\ref{fig:prob_sniping_geom_basics}) 
does not depend on the sniping probability. Indeed, 
a straightforward computation shows: 

\begin{equation}
s^{**} = \frac{A}{A+B} = \frac{m}{m - \overline{\alpha} q}  
\label{eq:s_double_star}
\end{equation}
This means that geometrically speaking, probabilistic 
sniping makes the bandit's utility function pivot 
about the intersection point  $s^{**}$ towards 
the horizontal axis  
(see Fig.~\ref{fig:prob_sniping_geom_basics}). 

\begin{verbatim}
For more details, see:
KPD_supplementary_material_1.ipynb(section.2.1)
\end{verbatim}

\subsubsection{Expected utility for market maker} 

The computation of the expected utility $\E U_M$ 
proceeds along similar lines but is somewhat more 
involved.  Again, we start by conditioning on the 
trigger event (Tr) which can either be the arrival 
of news (N, prob = $\beta$ ) or a liquidity trader 
(LT, prob = $1-\beta$): 
$$ \E U_M = \E (U_M\given Tr = N) \beta  + 
 \E (U_M\given Tr = LT) (1-\beta)
$$
The second term (trigger event is the 
{\it arrival of a liquidity trader}) is relatively 
simple as no sniping is involved. 
A straightforward computation yields: 

\begin{equation}
    \E (U_M \given Tr = LT)  = (1+\overline{\mu})s - 
    \overline{\alpha} q (1- s)
\end{equation}

In case the trigger event is {\it news},  
the market maker's payoff 
depends on whether or not he loses the 
ensuing race and therefore gets sniped.  
Using the notation introduced above we conclude: 
\begin{eqnarray*}
 \E(U_M\given Tr = N) &=& \E(U_M \given Tr = N \,\, \& \,\,
 \mbox{MM loses race})\,  h(p) \\
&+& \E(U_M \given Tr = N \,\,\& \,\, \mbox{MM wins race} ) 
 (1-h(p))
\end{eqnarray*}
Again, a straightforward but tedious calculation 
shows that the market maker's utility in case 
he gets sniped equals: 

\begin{eqnarray}
    \E(U_M \given Tr = N \,\, \& \,\,
 \mbox{MM loses race}) & = & (2\overline{\mu}-\overline{\alpha} q) s 
 -\gamma(1-\overline{\mu})(1-s)
\end{eqnarray}
Similarly: 
\begin{equation}
    \E(U_M \given Tr = N \,\, \& \,\,
 \mbox{MM wins race})= ((\gamma + 1)s -\gamma+1) \, \overline{\mu}
\end{equation}

From the expressions above we can constitute the linear 
function $\E U_M(s)$  but in the 
remainder of the paper we only need explicit expressions 
for the endpoints: 

\begin{equation}
    \begin{array}{rclcl}
        C \equiv C(p,\gamma) & :=  & \E U_M(s=0) &= 
        & -\left\{ q\overline{\theta}+\beta(m\gamma - \overline{\mu} q)  h(p)\right\} \\
        D \equiv D(p,\gamma) & :=  & \E U_M(s=1) &=& 
        (1+\overline{\mu})- \beta(m + \overline{\alpha} q h(p))
    \end{array}
    \label{eq:C_and_D_fion_of_p_and_gamma}
\end{equation}
 \begin{verbatim}
See also: KPD_supplementary_material_1.ipynb (section.2.2)
\end{verbatim}

\subsubsection{Indifference spread and utility for probabilistic sniping}
\label{sct:equilibrium_prob_sniping}

From the discussion above we know that for 
probabilistic sniping the end-points 
of the linear utility functions are replaced by 
their probabilistic counterparts: 

$$ A \rightarrow  A(p), \quad B \rightarrow B(p), 
\quad C \rightarrow C(p),\quad \mbox{and}\quad 
D \rightarrow D(p).$$
Using the same logic as before we conclude that the point of 
indifference for probabilistic sniping is given by  (cf. eqs.~\ref{eq:u_star_sure}): 
%
%
\begin{eqnarray}
s^*(p) & = & \frac{A(p)-C(p)}{(A(p)-C(p))+(D(p)-B(p))}
    \label{eq:s_star_p} \\
   u^*(p)  & = & \frac{A(p)\,D(p)-B(p)\,C(p)}{(A(p)-C(p))+(D(p)-B(p))}
    \label{eq:u_star_p}    
\end{eqnarray}
For ease of reference, we will denote the common denominator by 
\begin{equation}
 Q(p) :=  A(p)-C(p)+D(p)-B(p), 
\label{def:def_Q} 
 \end{equation}
and introduce
\begin{equation}
 N(p) := A(p)\,D(p)-B(p)\,C(p). 
 \label{eq:def_N}
\end{equation}
Hence, we arrive at the useful shorthand
\begin{equation}
    u^*(p)  = \frac{N(p)}{Q(p)}.
\label{eq:u_star_p_short}    
\end{equation}

 \begin{verbatim}
See also: KPD_supplementary_material_1.ipynb (section.4)
\end{verbatim}

\subsubsection{Impact of risk aversion and probabilistic 
sniping on utilities}  
\label{sct:impact_risk_aversion}

Since the point of indifference is completely determined 
by the utility endpoints (A, B, C and D), it is helpful 
to make their dependence on the parameters $\gamma$ 
(risk aversion) and $p$ (individual sniping probability) 
explicit. To this end we hark back to 
eqs.~(\ref{eq:A_and_B_fion_of_p_and_gamma}) and 
 (\ref{eq:C_and_D_fion_of_p_and_gamma}), from which we can 
compute: 

\begin{equation}
\begin{array}{lcl}
{\displaystyle \frac{dA}{d\gamma}} = 0, & \quad\quad\quad  &
{\displaystyle\frac{dB}{d\gamma}} = -\overline{\alpha} \beta p g(p) <0,  \\[3ex]
{\displaystyle\frac{dC}{d\gamma}} 
= - (\overline{\theta} + \beta(1-2\overline{\mu}) h(p)) < 0, 
& \quad\quad & 
{\displaystyle\frac{dD}{d\gamma}} = -\overline{\alpha} \beta h(p) <0.
\end{array}
\end{equation}
It therefore transpires 
that increasing risk aversion pushes 
the utility endpoints (except for A) down, dragging 
the utility $u^*$ along.  We will return 
to this observation in 
section~\ref{sct:transition_geom} 
where we will show how increasing risk-aversion will 
induce transitions in sniping behaviour. 

\begin{figure} [H]
    \centering
    \includegraphics[width=0.8\textwidth]{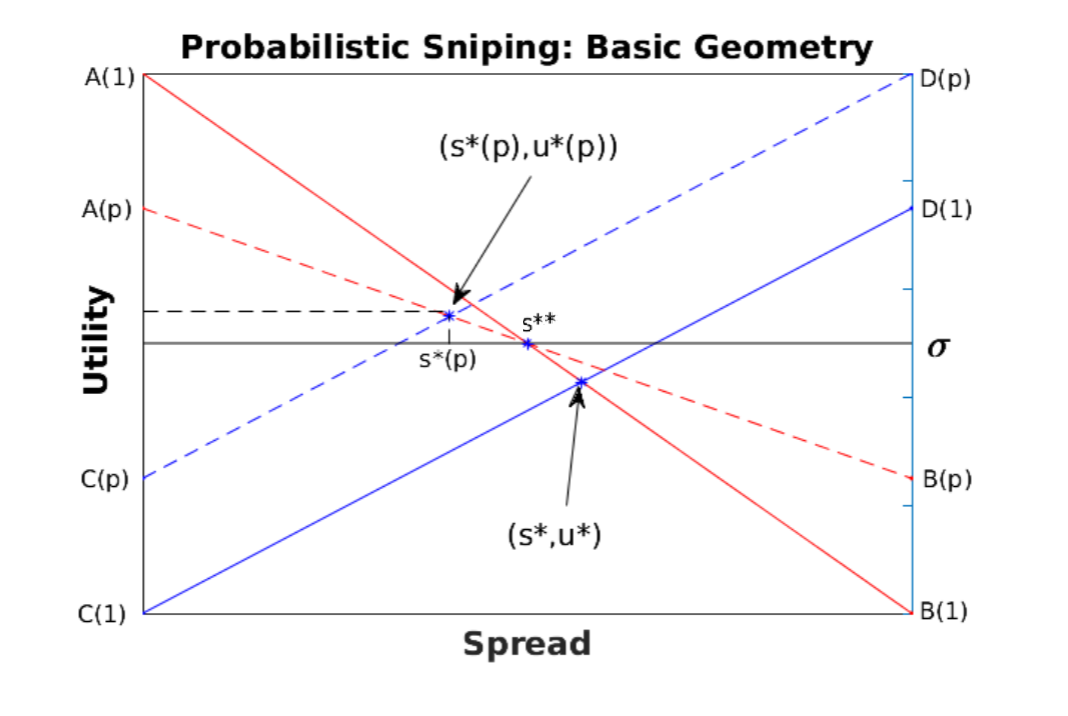}
    \caption{Basic geometry of probabilistic sniping: The solid lines (AB for the bandit, and CD for the market maker) 
    represent the expected utilities (as function of the spread $s$) in the case of sure sniping. The resulting 
    point of indifference 
    $(s^*,u^*)$  is non-playable since $u^*<0$. Reducing the 
    probability of sniping to  $p<1$ pushes the MM utility function 
    CD upwards (as the MM benefits from less sniping), while 
    the bandit's utility pivots about its zero crossing. 
    As a consequence, the point of indifference moves to 
    the new intersection point
     $(s^*(p),u^*(p))$ which is now playable as $u^*(p)>0$. }
    \label{fig:prob_sniping_geom_basics}
\end{figure}

The dependence of the utilities on the sniping probability $p$ can be  classified in a similar 
fashion:  
\begin{equation}
\begin{array}{lcl}
{\displaystyle\frac{dA}{dp}} = m \beta (g(p) + pg'(p)) >0, & \quad\quad\quad  & 
{\displaystyle \frac{dB}{dp}} = -\overline{\alpha} q \beta (g(p) + pg'(p))<0 \\[3ex]  
{\displaystyle\frac{dC}{dp}} = - m\beta h'(p) < 0,  & \quad &
{\displaystyle\frac{dD}{dp}} 
= -\overline{\alpha} \beta q h'(p) < 0.
\end{array}
\label{eq:derivs_endpoints_wrt_p}
\end{equation}
Notice that since $g(p) + pg'(p) = (p g(p))' = h'(p)/(H-1)>0$, 
it follows that 
(see Fig~\ref{fig:prob_sniping_geom_basics})
probabilistic sniping (i.e. reducing $p$) will 
pivot the bandit's utility towards smaller values 
about the fixed intersection point $s^{**}$.  
Similarly, since both $dC/dp<0, dD/dp<0$ the utility 
for the market maker (unsurprisingly) 
shifts upwards when 
$p$ is lowered and the 
bandits snipe only occasionally.  
 \begin{verbatim}
See also: KPD_supplementary_material_1.ipynb (section.3)
\end{verbatim}

\subsection{When is probabilistic sniping advantageous? } 
\label{sct:prob_sniping_advantageous}

\subsubsection{Geometric interpretation: Probabilistic sniping shifts point of indifference}
 
\begin{figure}[htbp]
    \centering
    \includegraphics[width=.45\textwidth]{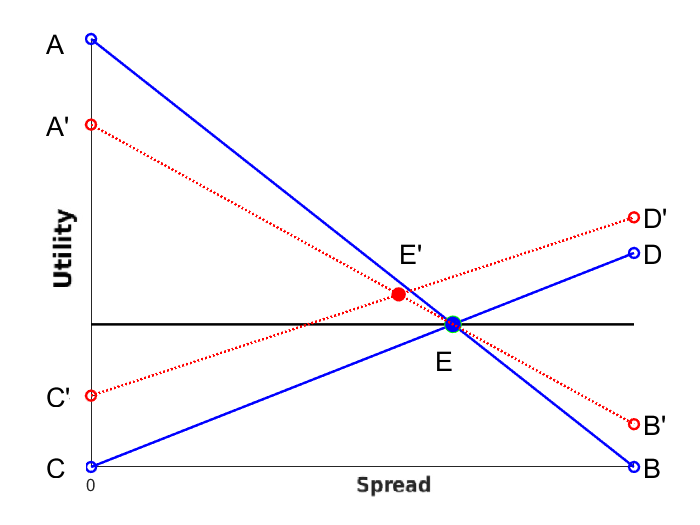}
    \includegraphics[width=.45\textwidth]{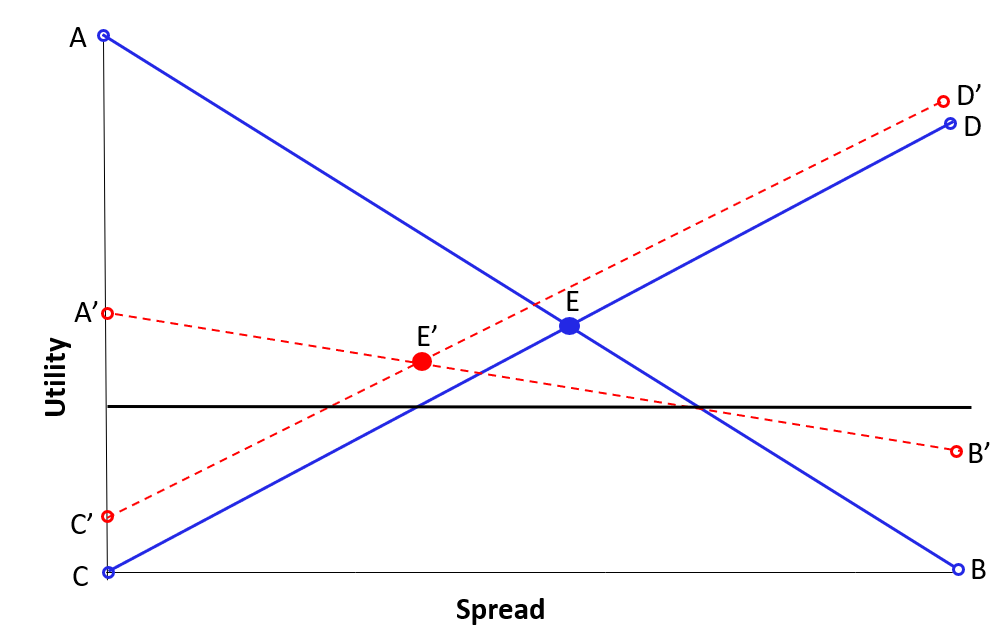}
    
        \caption{LEFT: Geometric representation that illustrates why 
    probabilistic sniping is advantageous when $u^*=0$. 
    The blue lines AB (bandit) and CD (market maker) 
    represent the expected utilities under sure sniping 
    yielding a point of indifference E corresponding to 
    zero utility $u^*(p=1) = 0$. Probabilistic 
    sniping (with sniping probability $p<1$) shift these 
    utilities to A'B' and C'D' respectively, resulting 
    in a new point of indifference E' that corresponds to 
    strictly better utility $u^*(p<1)>0$. 
    RIGHT: Configuration in which probabilistic sniping 
    results in lower expected utility and sure sniping 
    is to be preferred. 
    See main text 
    for more information. 
}
    \label{fig:prob_sniping_geom}
\end{figure}

Is there an advantage in switching from sure sniping to 
probabilistic sniping? Obviously, less sniping is advantageous 
for the market maker as his pay-off is reduced by sniping. 
In section \ref{sct:rational_prob_sniping} we argued that less sniping might also 
protect bandits from acquiring assets whose values 
are unstable (e.g. due to subsequent emergence of  contradictory news). 
Some further  intuition 
is gleaned from Fig.~\ref{fig:prob_sniping_geom}, which illustrates 
the difference between   
sure and probabilistic sniping.  More specifically, 
let us assume that sure sniping results in an equilibrium $E$
such that $u^* := u(s^*) = 0$.  Sure sniping ($p = 1$) is represented 
by the two blue lines (AB and CD, expected utility for bandit and market maker,  
respectively).  Changing to probabilistic sniping ($p<1$) results 
in two different lines ($A'B'$ and $C'D'$, respectively).  
Notice that $A'B'$ pivots about the equilibrium point E, while 
$C'D'$ shifts upwards (as less sniping means 
less risk for the 
market maker).  As a consequence, the new 
intersection point $E'$
will yield 
a (strictly) positive utility. 

Another way to 
put this is that the derivative $du^*/dp$ (evaluated at
$p=1$) is negative: smaller values for $p$ 
result in higher utility 
(for a concrete, numerical example, 
see Fig~\ref{fig:prob_sniping_example_1} below). 
In section \ref{sct:transition_geom} below,   
we will recast this geometric insight into an algebraic expression, but first 
we will show that this new equilibrium 
is not a Nash equilibrium for the one-shot MZ$^1$.

\begin{figure}[htbp]
    \centering
       \includegraphics[width=.75\textwidth,height=7.5cm]{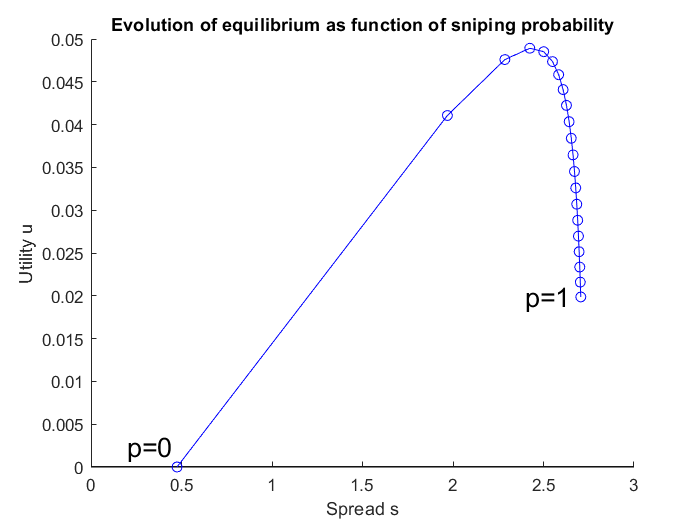}
    \caption{
    An example where probabilistic sniping is advantageous even though  sure sniping results in a strictly positive utility. 
    In this example, 
    we depict the evolution of the point of indifference 
    as a function of the sniping probability $p$, i.e. 
    $(s^*(p),u^*(p))$, where  $p$ starts at $p=1$ 
    and is decremented with steps of 0.05 to $p=0$.  Notice how the 
    expected utility rises from $0.02 $ for sure sniping ($p=1$) 
    to a maximum of 0.05 for $p \approx 0.15$, before 
    dropping to zero when there is no sniping ($p=0$). 
    }
    \label{fig:prob_sniping_example_1}
\end{figure}

\subsubsection{Does $(s^*(p),u^*(p))$ constitute a Nash equilibrium 
of the MZ$^1$ game?}
\label{sct:nash_equilibrium_Mz_game}
One of the reasons why 
for $p<1$ 
the new intersection point 
$(s^*(p),u^*(p))$  
is not considered in the MZ paper (even if  $u^*(p)>0$), is 
that it does not constitute a Nash equilibrium for the 
(single-shot) MZ game.   
To see this, 
it suffices to realise 
that when $p<1$, a  bandit  
will always benefit from unilaterally deviating 
to sure sniping (assuming $u^*(p) \equiv U_B(s^*(p))>0$).  
As a consequence,  these intersection points 
are uninteresting 
in terms of the single-shot version of the MZ game.  

However, 
the reason that probabilistic sniping (and the corresponding 
role of $s^*(p)$ and $u^*(p)$) is  of interest nonetheless, 
is that 
things change when we will consider (in section~\ref{sct:app_repeated_games})
the infinite horizon repeated game version (MZ$^\infty$).    
The promise of higher pay-offs means that 
even purely selfish agents are  incentivised to collaborate if 
they expect this cooperation to be profitable. 
This new equilibrium is stabilized by the  
implicit threat that, if anyone breaks the tacit plan, 
opponents will do the same and 
opportunities will instantly evaporate.  Working out whether cooperation 
is the rational choice involves estimating
what is most favourable: the exceptional but one-off payoff one stands to gain 
from defecting, or the smaller but accumulating payoffs that result from 
continuing collaboration  that over time, will eclipse the former.  
For more details, we refer to section \ref{sct:app_repeated_games}.

\section{Characterising the transitions 
in sniping behaviour}
\label{sct:transition_geom}

In this  section we use  
a geometric argument to 
deduce the general conditions 
under which, first, the transition from sure to 
    probabilistic sniping occurs, and second, 
probabilistic sniping ceases to be profitable 
(i.e. transition from probabilistic to no sniping). 
We will then show how these 
general conditions can be related 
to risk aversion $\gamma$ and how they give rise to two thresholds 
($\overline\gamma_K$ and $\overline\gamma_L$) that 
govern these transitions. 

\subsection{Transition from pure to probabilistic sniping}
\label{sct:from_pure_to_prob}

\paragraph{Geometric interpretation}
As argued above, 
the equilibrium  $(s^*,u^*)$ can be seen as  
 a function of the sniping probability $p$, i.e. 
$(s^*(p), \,u^*(p)) $.  
To decide whether probabilistic sniping is advantageous, we need to 
 determine when the 
slope of the tangent to the curve $u^*(p)$ 
at $p=1$ changes sign (see Fig.~\ref{fig:prob_sniping_examples}):
\begin{equation} 
\left. \frac{du^*(p)}{dp} \right|_{p=1} > 0 \, 
\mbox{(sure sniping)}\quad \longrightarrow \,
\left. \frac{du^*(p)}{dp} \right|_{p=1} < 0 \, \mbox{(probabilistic sniping)} 
\label{eq:transition_pure_prob_general}
\end{equation}
Indeed, if the slope of this tangent is positive, it means 
that reducing $p$ from 1 (sure sniping) to a lower 
value $p<1$ decreases the value of $u^*(p)$.  Hence, 
sure sniping is better than probabilistic sniping. 
Conversely, if the slope of the tangent is negative, 
moving from sure to probabilistic sniping (i.e. 
reducing $p$) does improve $u^*(p)$ and therefore probabilistic sniping is 
to be preferred. 

Using the notation introduced in eq.~(\ref{eq:u_star_p_short})
 we know that 
 $u^*(p) = N(p)/Q(p)$, 
and hence 
\begin{equation} 
\left. \frac{du^*(p)}{dp} \right|_{p=1} =
\frac{N'(1)Q(1) - N(1)Q'(1)}{Q^2(1)}
\label{eq:dudp_at1}
\end{equation}
%
%
%
Hence the threshold is determined by the equation: 
\begin{equation}
    \left. \frac{du^*(p)}{dp} \right|_{p=1} = 0 
    \quad\quad \Longleftrightarrow \quad \quad 
    N'(1)Q(1) - N(1)Q'(1)= 0
    \label{eq:dN_dp_1}
\end{equation}
 \begin{verbatim}
See : KPD_supplementary_material_1.ipynb (section.5)
\end{verbatim}

\subsection{Transition from probabilistic to 
cessation of sniping (non-sniping)}
\label{sct:from_prob_to_non}

If the expected utility associated with the sure sniping 
equilibrium is sufficiently negative (i.e. $u^*(s^*,p=1)\ll 0$), 
even probabilistic sniping will not be able to 
turn this into profits. Geometrically, this 
corresponds to a situation in which the path  
traced by $(s^*(p), u^*(p))$ will never  
enter (strictly) positive territory before 
it hits the x-axis at $p=0$ 
(see Fig.~\ref{fig:prob_sniping_examples} bottom-right panel).

This transition can therefore be characterised by a 
condition which is analogous 
 to the characterisation 
of the transition 
{\it from pure to probabilistic 
sniping} in eq.(\ref{eq:transition_pure_prob_general}), 
this time however focusing  
on the tangent at $p=0$:
\begin{equation} 
\left. \frac{du^*(p)}{dp} \right|_{p=0} > 0 \,\,
\mbox{(probabilistic sniping)}
\quad \longrightarrow  \,\,
\left. \frac{du^*(p)}{dp} \right|_{p=0} < 0 \,\, \mbox{(no sniping)} 
\label{eq:transition_prob_non_general}
\end{equation}
This transition therefore occurs when 
\begin{equation}
    \left. \frac{du^*(p)}{dp} \right|_{p=0} = 0
    \label{eq:transition_prob_non_general_2}
\end{equation}
Expanding  
$$\left. \frac{du^*(p)}{dp} \right|_{p=0} = \frac{N'(0)Q(0)-N(0)Q'(0)}{Q^2(0)}$$
and using that  $N(0)=0$ while   $Q(0)=D-C>0$, we 
see that eq.~(\ref{eq:transition_prob_non_general_2}) 
simplifies to: 

\begin{equation} 
N'(0) = 0 .
\label{eq:N_prime_0}
\end{equation}
Expanding equation (\ref{eq:N_prime_0}):   
$$N'(0) = 
A'(0)D(0)+A(0)D'(0)-B'(0)C(0)-B(0)C'(0)
$$
and using 
$A(0) = B(0)=0$, 
condition (\ref{eq:N_prime_0})  
can be further simplified to: 
\begin{equation}
 N'(0) \equiv  A(0)D'(0) - B'(0) C(0)  = 0 .
 \label{eq:N_prime_0_bis}
\end{equation} 
 \begin{verbatim}
See : KPD_supplementary_material_1.ipynb (section.6)
\end{verbatim}

\subsection{Increasing risk aversion induces transitions}
\label{sct:role_of_gamma}
In section \ref{sct:impact_risk_aversion}  we showed how 
the endpoints of the utilities 
(except for A) are pushed down by increasing risk aversion $\gamma$, dragging 
the point of indifference $(s^*,u^*)$ down in their wake.  
Since $u^*$  decreases, there will be a value for $\gamma$ beyond  
which probabilistic sniping becomes more favourable than 
sure sniping. 
Increasing $\gamma$ even further will eventually result in 
a utility that is so low that even probabilistic sniping is 
no longer profitable.  
Hence we see that increasing risk aversion $\gamma$ is one way to 
induce the two transitions in sniping behaviour discussed 
above. In this section we will translate the 
general transition conditions (\ref{eq:dN_dp_1})  and  (\ref{eq:N_prime_0_bis}) into 
precise transition thresholds for $\gamma$.

\subsubsection{Threshold $\overline\gamma_K$: Transition from sure to probabilistic sniping}

Both $N$ and $Q$ depend on risk aversion $\gamma$ and expanding 
the transition condition 
eq.~(\ref{eq:dN_dp_1}) as a function of $\gamma$
yields a cubic polynomial  (denoted by $K(\gamma)$), 
for which we need to find a zero-crossing: 

\begin{equation}
    N'(1)Q(1)  - N(1)Q'(1) \equiv 
    \underbrace{K_3\gamma^3 + K_2\gamma^2 + K_1 \gamma 
    + K_0}_{:= K(\gamma) } = 0 
    \label{eq:K_cubic}
\end{equation}
The computation of the polynomial coefficients is cumbersome 
and uninteresting and is therefore relegated to the 
supplementary material.
\begin{verbatim}
   See: KPD_Supplementary_material_1.ipynb (section.5.1)
\end{verbatim}

In terms of this notation, the threshold 
condition in eq.~(\ref{eq:dN_dp_1}) amounts to finding the 
appropriate zero-crossing $K(\gamma) = 0$. 
Recall that a cubic equation 
has at least one solution (zero-crossing).  However, we need 
to make sure that this solution satisfies $\gamma \geq 1$ 
and can therefore  be interpreted as a risk aversion factor. 
This is proven in the following theorem in which 
we also define the transition threshold:

\begin{thm}[Risk aversion threshold for transition from pure to probabilistic sniping] 
\label{thm:gamma_K}
The cubic polynomial $K(\gamma)$ specified in 
 eq.~(\ref{eq:K_cubic}) has a unique zero-crossing 
greater than 1.  We will denote this unique zero-crossing 
by $\overline{\gamma}_K \geq 1$ as it is the 
threshold that governs the transition from sure to probabilistic 
sniping. 
\end{thm}
The proof of this proposition can be found in appendix~\ref{appx:trans_pure_prob}.


\begin{figure*}[htbp]
\centering
\begin{multicols}{2}
   \includegraphics[width=0.5\textwidth]{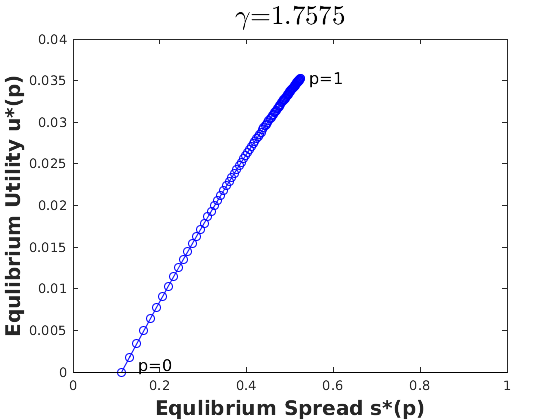}\par 
    \includegraphics[width=0.5\textwidth]{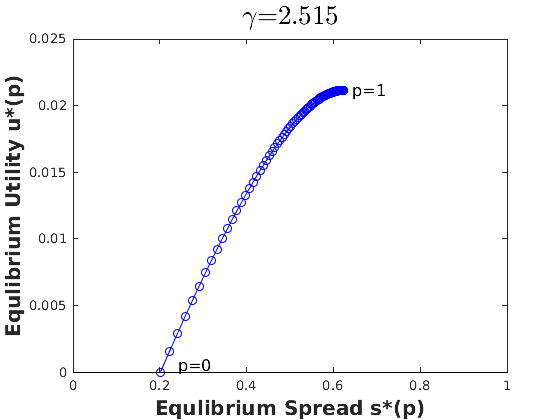}\par 
    \end{multicols}
\begin{multicols}{2}
\includegraphics[width=0.5\textwidth]{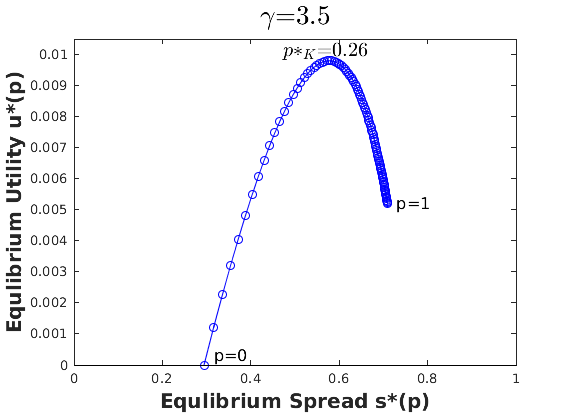}\par
    \includegraphics[width=0.5\textwidth]{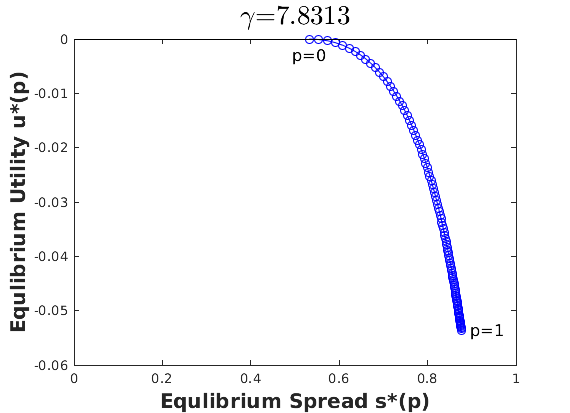}\par
\end{multicols}
\caption{Examples  of transition thresholds  in sniping 
behaviour. In the above example we fix the parameters 
     $\alpha = .45, \,\mu = .5, 
     \, \delta = .5, \,
    H = 5$.  From these values we can compute 
    the threshold values $\overline\gamma_K= 2.515$ 
    and $\overline\gamma_L = 7.831$ 
(for a definition of $\overline\gamma_K$  and 
$\overline\gamma_L$ see section \ref{sct:transition_geom}). 
  Next we set $\gamma $ to four 
    different values:  $\gamma = 1.7575$ (slightly  risk-averse,  sure sniping), 
    $\gamma =\overline{\gamma}_K=2.5150 $ (risk averse, knife edge),  $\gamma = 3.5$ (probabilistic 
    sniping), and finally, $\gamma = \overline\gamma_L= 7.8313$ (strongly risk-averse, no more sniping). 
    Notice that in the bottom left picture 
    probabilistic sniping yields a better utility, even though 
    the utility for 
    sure sniping is positive!
 }
    \label{fig:prob_sniping_examples}
\end{figure*}
 \begin{verbatim}
See also: KPD_supplementary_material_numeric.ipynb (section.3.4)
\end{verbatim}

\subsubsection{Threshold $\overline\gamma_L$: Transition from probabilistic sniping to non-sniping}

A similar approach can be taken to 
quantify how a further increase in risk aversion $\gamma$ gives rise 
to the threshold $\overline\gamma_L$ beyond which even probabilistic sniping 
is no longer profitable, i.e. both sure 
and probabilistic sniping results in 
negative utility $u^*(s^*,p)<0, \quad 
(\forall \, \,0\leq p\leq 1)$.

In order to determine this threshold value, 
we expand transition condition eq.(\ref{eq:N_prime_0_bis}) as a function 
of $\gamma$ and obtain a quadratic polynomial (denoted by $L(\gamma)$) 
for which a zero-crossing needs to be found:

\begin{equation}
    N'(0) = 0 \quad \quad \Longleftrightarrow 
    \quad \quad 
    L(\gamma) := L_2(\gamma-1)^2 + L_0 = 0, 
    \label{eq:L_cubic}
\end{equation}
which can easily be solved explicitly. 
We can summarize all of these observations in the following 
proposition.

\begin{thm}[Risk aversion threshold for transition from probabilistic to non-sniping] 
There is a unique threshold $\overline\gamma_L$ 
for risk aversion $\gamma$ that 
governs the transition from probabilistic to no sniping.  Specifically:  
\begin{equation}
\overline\gamma_L =  
1 + \sqrt{\frac{(1-\overline{\mu}) Z}{\overline{\alpha}\, \overline{\theta} }} \quad \quad 
\mbox{where }\quad   Z = 1+\overline{\mu} - \beta(1-\overline{\mu}). 
\label{eq:def_gamma_L}
\end{equation}
\label{thm:gamma_L}
\end{thm}
The proof of this proposition is straightforward 
and can be found in Appendix section~\ref{appx:trans_prob_non}.
 \begin{verbatim}
See: KPD_supplementary_material_1.ipynb (section.6)

\end{verbatim}

\subsection{Optimal sniping probability $p^*_K$ }
\label{sct:p_star_K}

\begin{figure}[htbp]
    \centering
    \includegraphics[width=0.9\textwidth,height=7.5cm]{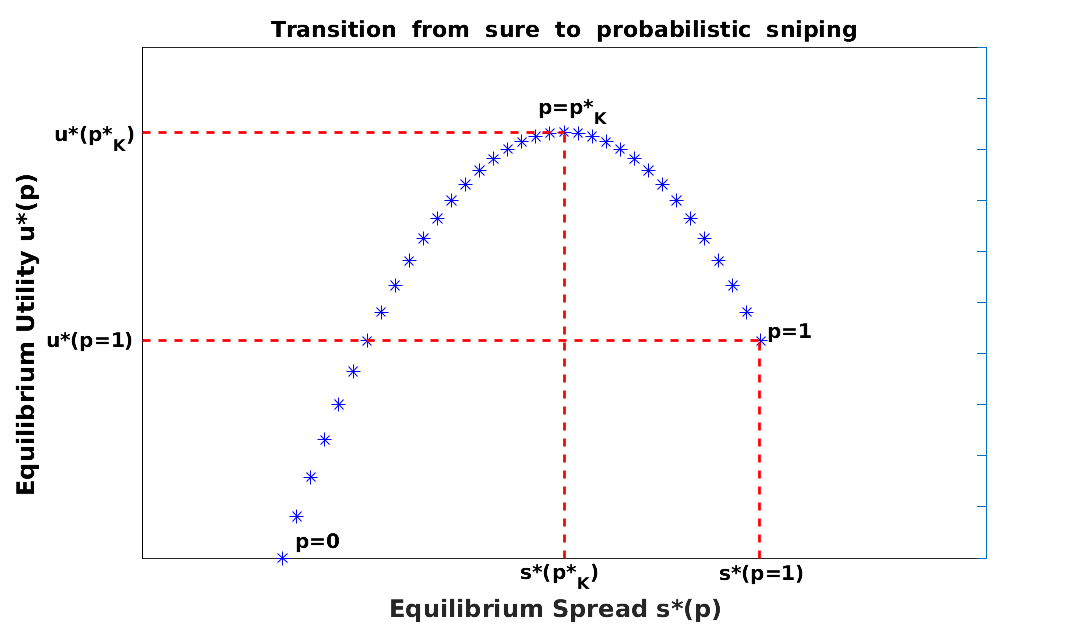}
    \caption{Definition of $p_K^*$:  When probabilistic sniping 
    is advantageous (i.e. 
    $\overline\gamma_K <\gamma  < \overline\gamma_L$) 
    reducing the sniping probability $p$ shifts the equilibrium 
    position $(s^*(p),u^*(p))$ along a concave arc, the top of 
    which corresponds to the optimal sniping probability $p^*_K$. }
    \label{fig:p_star_K_definition}
\end{figure}

\begin{figure}[htbp]
    \centering
    \includegraphics[scale=0.6]
    {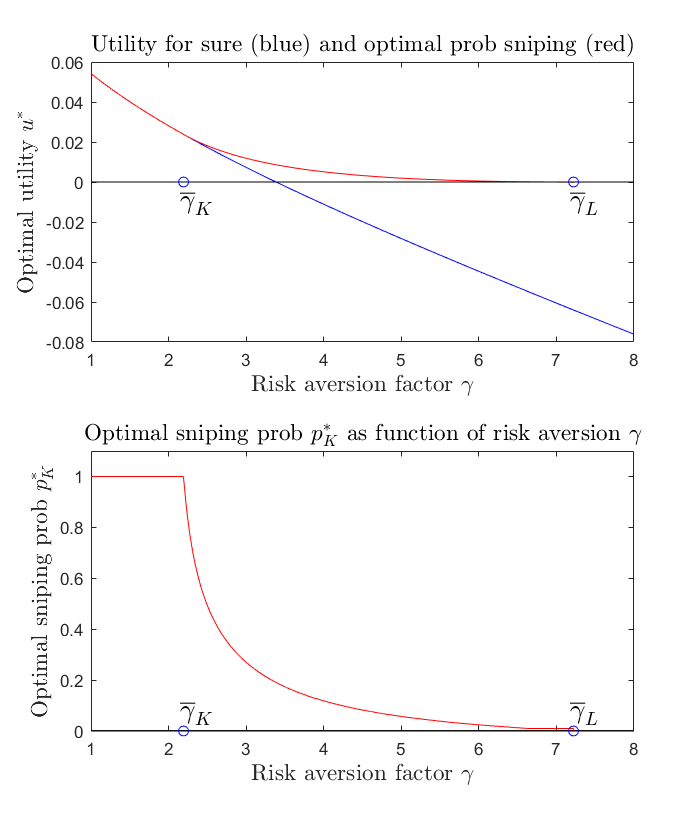}
    \caption{TOP: The figure shows the equilibrium utility spread as a function of risk aversion in both sure sniping 
    ($u^*_K(p=1)$ blue line) and 
    optimal 
    probabilistic sniping ($u^*_K = u(p=p^*_K)$, red line). 
    BOTTOM: corresponding optimal sniping probability $p^*_K$, which drops from $p^*_K=1$ (sure sniping) 
    when $\gamma \leq \overline\gamma_K$ to 0  (no sniping)
    when $\gamma \geq \overline\gamma_L$.}
    \label{fig:u_star_K_as_function_of_gamma}
\end{figure}

In the preceding sections we have shown that 
increasing risk aversion ($\gamma >1$) 
creates multiple sniping regimes for the 
bandits: 

\begin{itemize}
    \item $1\leq \gamma < \overline\gamma_K$: 
    Sure sniping is most profitable; 
     This situation corresponds to 
    the top left panel in 
    Fig.~\ref{fig:prob_sniping_examples}.
    \item $ \overline\gamma_K\leq \gamma < \overline\gamma_L$:  Probabilistic sniping 
    results in higher utilities; in fact, there 
    is an optimal sniping probability $p^*_K$ (see 
    below) which results in a maximal (positive) 
    utility. This situation corresponds to 
    the bottom left panel in 
    Fig.~\ref{fig:prob_sniping_examples}.
    \item $\gamma \geq \overline\gamma_L$: No sniping 
    (resulting in zero utility) is the best 
    option.  This situation corresponds to the 
    bottom right panel in 
    Fig.~\ref{fig:prob_sniping_examples}.
\end{itemize}
The situation 
in the second case 
$\overline\gamma_K < \gamma < \overline\gamma_L$
is schematically illustrated 
in Fig.~\ref{fig:p_star_K_definition}. 
It follows that there is an optimal 
sniping probability that yields the largest 
utility: 
\begin{equation}
    p^*_K := \arg\max_p u^*(p)
    \label{eq:optimal_p}
\end{equation}
We introduce the 
following notation for the corresponding utility $u^*_K$ 
and spread $s^*_K$.
Since this is a unique optimal point it provides 
a natural {\bf focal point} for the game~\cite{Schelling}. 
In addition, 
Fig.~\ref{fig:u_star_K_as_function_of_gamma}  compares the expected utilities 
under both sure and (optimal) probabilistic sniping 
as a function of risk aversion $\gamma$, showing 
clearly that profitability can be sustained 
for higher values of $\gamma$.

\begin{verbatim}
    See: u_depend_on_gamma_p_v2.m
\end{verbatim}

\section{Optimal probabilistic sniping as  SPE in  
repeated MZ\texorpdfstring{$^\infty$}{Lg} game}
\label{sct:app_repeated_games}

In the previous section we have shown that 
if the risk aversion factor $\gamma$ satisfies 
$\overline\gamma_K < \gamma < \overline\gamma_L$, all bandits  
are better off if they implement probabilistic sniping, i.e. 
engage in racing with probability $p^*_K$ only. However, this is 
not a Nash equilibrium for the stage game, as bandits 
will be tempted to snipe more frequently. This is 
a situation reminiscent of  the classical prisoner's dilemma  (PD)
where the Nash equilibrium forces the players into 
a strategy that results in the worst 
social welfare possible~\cite{Holt2004}. 

However, it is well-known that if we repeat the prisoner's dilemma 
an indefinite number of times, it becomes possible to devise 
strategies that do yield the optimal utility. 
This reversal of fortune is due to the fact that agents 
can build up reputations that encourage collaboration 
that is guaranteed to be profitable for all parties. 
Exactly the same rationale will also play out in the 
(infinite horizon) repeated 
MZ-game (MZ$^\infty$).

\subsection{The MZ stage game as a sequential game} 
\label{sct:mz_sequential}

\paragraph{Game Tree}

Before we explore the strategies that are 
available in the (indefinitely) repeated MZ game
(a.k.a. MZ$^\infty$), 
we need to pause for a minute and clarify 
the nature of the strategies that are available 
to the players in the single-shot stage game 
(a.k.a. MZ$^1$). 
To this end, it is helpful to realise that 
the stage game 
involves different 
interdependent steps and can therefore 
be seen as a sequential game
involving a (possibly) random move by 
``nature''.   Below, we give an explicit 
representation of the decision tree for  the 
game and clarify a 
subtle but important  
point in the interpretation of probabilistic sniping. 

\begin{figure}[htbp]
    \centering
    \includegraphics[width=0.8\textwidth]{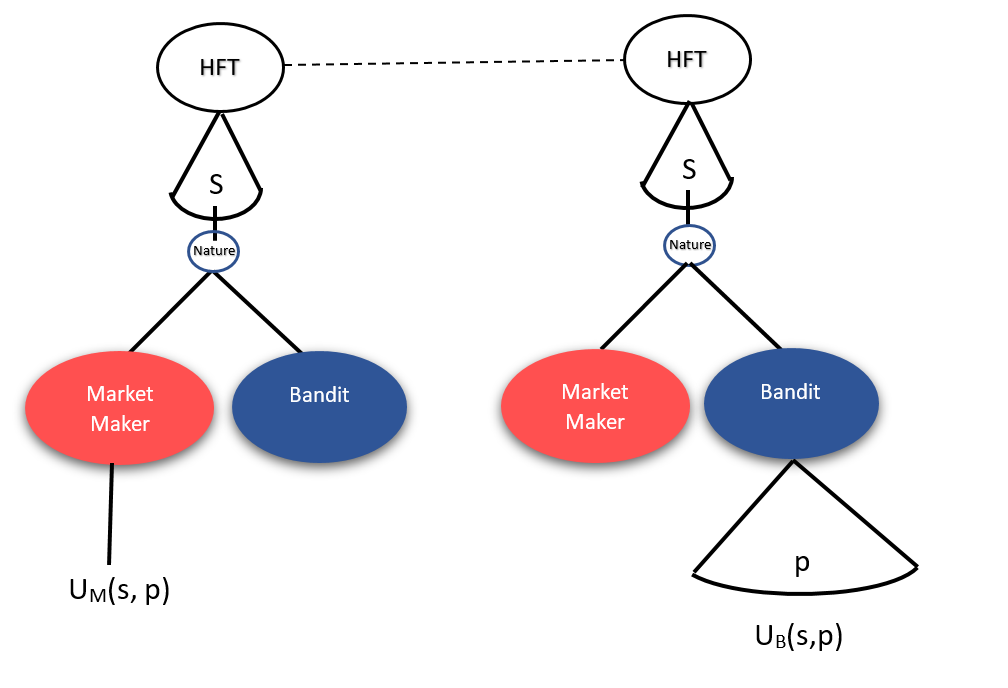}
    \caption{The MZ game as a sequential game against nature. 
    Notice that in the last decision point, the MM is uncertain 
    about his utility $u_M(s,p)$ as it depends 
    on the bandit's sniping probability which at the moment 
    of decision is unknown to him!}
    \label{fig:mz_sequential_game}
\end{figure}

In section~\ref{sct:mz-game_ingredients}, 
we already outlined the detailed chronology of 
the various defining 
events in the game.   For the interpretation of 
the MZ game as a sequential game 
it is also helpful to highlight the various steps
for an individual HFT 
in the 
corresponding decision tree  
(also see  Fig.~\ref{fig:mz_sequential_game}): 
\begin{enumerate}
    \item HFT (at time $t=-1$) has to choose a spread $s$ 
    from the continuous action space $0\leq s \leq \sigma$. 
    Alternatively,  he can refrain from playing altogether 
    (NULL action resulting in zero utility).
    \item Nature (at time $t = -1/2$) selects one of the HFTs as MM 
    (using the rule specified earlier 
    in section \ref{sct:mz-game_ingredients}), relegating 
    the others to the role of bandit; 
    \item At $t=0$, each HFT knows its type (market maker 
    or bandit) and 
    can therefore make his reaction to the trigger event 
    (news or LT arrival) contingent on his type. 
    This results in a new bifurcation of the decision 
    tree: 
   
    \begin{itemize}
        
        \item if he's bandit:  he has to choose between 
        quitting the game (NULL-action with utility 0), or 
        some form of sniping: {\it sure} $p=1$, 
        {\it no} ($p=0$) or {\it probabilistic} sniping ($0<p<1$)
         resulting in a utility $u_B(s,p)$.
        \item if he's MM:  he either proceeds and the game ends with utility $u_M(s,p)$ (which 
        actually also depends on $p$), {\it or} he withdraws from the game (NULL with utility 0): 
        
    \end{itemize}
\end{enumerate}

\paragraph{Actions and Strategies}  

Disregarding the trivial NULL action, 
what  actions and strategies are available 
to the players? 
At the first decision point, 
it is clear that each HFT can choose a spread $s$, 
which corresponds to a {\bf  pure but continuous action. }

\begin{figure}[htbp]
    \centering
    \includegraphics[width=0.9\textwidth]{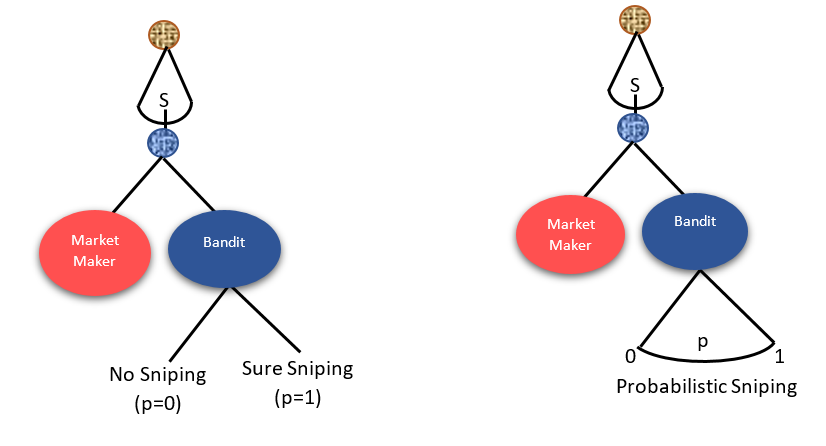}
    \caption{Interpretation of the sniping probability.  {\bf Left: } 
    In the MZ paper, the pure actions available to the bandit 
    are {\it sniping} and {\it not sniping}.  Mixed or (probabilistic) 
    sniping is therefore only possible when the bandit is indifferent 
    between the pure actions. {\bf Right:}  In this paper, we 
    interpret the sniping probability as a {\it pure 
    but continuous} action 
    that the bandit can take. In this interpretation the bandit need 
    not be indifferent between the extremes in order to adopt a 
    probabilistic strategy. }
    \label{fig:mz_seq_game_prob_sniping_interpretation}
\end{figure}

At the second HFT decision point, the HFTs know their 
type. The strategy for the market maker is straightforward: 
if the trigger event is the publication of a {\it news item}, 
the intrinsic value of the asset jumps and he will race 
to cancel his stale quotes.   The choice of strategies for  
the other HFTs (bandits) is more subtle.  Indeed, 
in the MZ paper~\cite{mz2017}, the authors describe 
    mixed sniping  as randomly mixing between the 
    {\bf pure and discrete strategies} of {\it  sniping} and  
    and {\it not sniping}.  As a consequence, mixed sniping 
    will only feature as a Nash equilibrium when the bandits 
    are indifferent between sniping and not sniping! 
    In our interpretation, we think of each bandit 
    selecting a 
    sniping probability $p$ as him choosing a {\it pure strategy} 
    from a {\bf continuous 
    action space.}  This allows us to introduce 
    probabilistic sniping in situations 
    where the expected utilities 
    at the extremes (i.e. $p = 0$ or 1) are not necessarily the 
    same.

\subsection{Subgame-Perfect Equilibria (SPE) in repeated games}
\label{sct:Subgame-Perfect Equilibria (SPE)}
\subsubsection{The emergence of collaboration in repeated games}  
\label{sct:collaboration in repeated game}

In the sections above we identified the risk 
aversion threshold 
$\overline\gamma_K $ above which probabilistic sniping has 
the potential of yielding better utilities for both the 
market maker and the bandits.  
When $\gamma > \overline\gamma_K $  there is a corresponding 
optimal sniping probability $p^*_K$  that yields the most 
favourable outcome for all parties. 
The problem with this strategy is 
that it does not constitute 
a Nash equilibrium for the single-shot MZ$^1$ game 
as bandits will be tempted to snipe more often than 
is allowed 
(see section~\ref{sct:prob_sniping_advantageous}). 
However, things are different when we consider the 
infinite horizon repeated game version (MZ$^\infty$).    
Indeed in this case, we can invoke the 
so-called 
{\it folk theorems} that show 
that any equilibrium that is strictly better than a 
Nash equilibrium in the single-shot (stage) game 
gives rise to a new Nash equilibrium (NE) in the 
repeated game (at least, under the mild assumption 
that all the players are sufficiently patient) \cite{Abreu}. 

However, for this to become a viable strategy 
we need to make the threat of rescinding the 
 cooperation implicit in probabilistic sniping,  
credible by devising a method to detect non-compliance 
among the other agents.   This is non-trivial since even deceptive agents would still advertise spread 
$s^*_K$, as spreads are publicly observable, 
and deviating from this would immediately flag their 
non-compliance and 
trigger a collapse of the (implicit) agreement.  
Deceptive agents would therefore stick to  spread $s^*_K$ 
but snipe for sure, rather than with probability $p^*_K$.  
However, the sniping behaviour of  agents is hidden to 
the other agents as  it is not publicly known who entered the 
race and who won it.  An agent only has access to 
its own utility (aggregated over many repetitions of the 
stage game).  So each agent will have to 
base his decision (on whether or not to continue with the 
game) on the impact that sniping by the deceptive agents 
has on his own utility.


\subsubsection{Impact of deceptive sniping}

When $\overline\gamma_K < \gamma < \overline\gamma_L$ 
and all agents snipe with optimal probability 
$p^*_K$ they stand to earn $u^*_K > 0$ (on average). 
If,  however, there are rogue agents that engage in 
sniping at every possible occasion, then the utility for the 
trustworthy agents will go down for two reasons: 

\begin{enumerate}
    \item as market makers they will suffer from more intense 
    sniping; 
    \item as bandits they will have to race against  
    an (on average) larger group of competing bandits.
\end{enumerate}
The combination of these two effects will affect their 
outcomes in various but related ways:  It will 
reduce the number of times they succeed in winning the race, 
and this in turn will  
push down their aggregated utility.  

\begin{figure}[htbp]
    \centering
    \includegraphics[height=7.5cm,width=\textwidth]{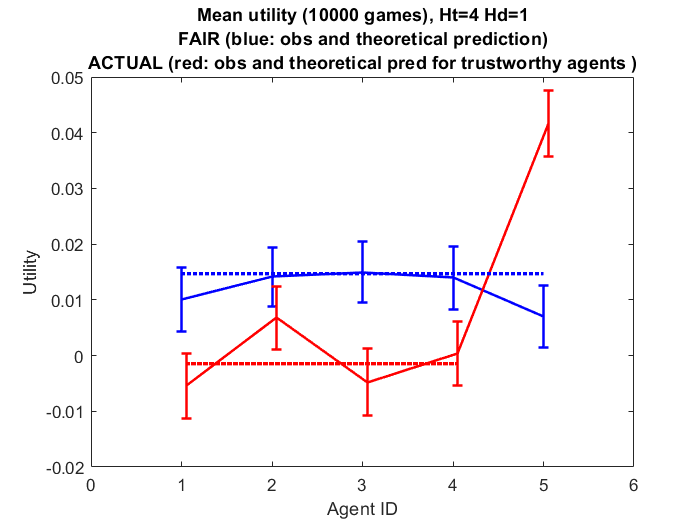}
    \caption{Mean utility for 
    agent group of $H_t =  4$ trustworthy 
    agents (agents 1 through 4) and $H_d = 1$ deceptive 
    agent (agent 5).  
    The mean and error-bars ($\pm 1$ standard error) 
    show the results 
    for a  simulation of a large number of 
    stage games, while the dotted lines 
    indicate the theoretically predicted value. 
    When everyone adheres 
    to probabilistic sniping,  the  mean is the 
    same for all agents.  When the deceptive agent 
    snipes for sure, the  theoretical mean (dotted red line) corresponds 
    to the expected utility for the trustworthy 
    agents (1 through 4). 
    The deceptive agent (agent 5) is clearly 
    able to increase his gains at the expense of the 
    trustworthy agents.}
    \label{fig:mean_utility_fair_vs_actual_3}
\end{figure}

This is illustrated in Fig.~\ref{fig:mean_utility_fair_vs_actual_3} 
where the blue error-bars show the average 
utility (averaged over 10000 
simulations of 
the stage game) when {\it all} agents snipe probabilistically (with probability $p^*_K$).  
These results are clearly consistent with 
the theoretically predicted value 
$u^*_K = 0.015$ (blue dotted line).  
However, things change 
significantly when there is even a 
single devious agent (HFT 5 in the figure) 
that snipes for sure 
(red error-bars). The utility for this deceptive 
agent more than doubles while the expected 
utility  for the 
trustworthy agents (1~through~4) dips below 
zero ($u = -0.0016$, red dotted line). 
\begin{verbatim}
See:KPD_supplementary_material_game_simulations.ipynb (section 7.2)
\end{verbatim}

 Observing this  unexpected downturn 
in utility an agent can suspect (some of) the other agents are not complying, and consequently decide to 
renege on his earlier commitment to the probabilistic sniping strategy. To turn this intuition into 
a reliable  operational strategy, we need to quantify 
the effect that deceptive agents have on the expected 
utility of the trustworthy agents.  This is taken up 
in the next section. 

\subsection{Expected utility in the presence of 
deceptive agents}

\subsubsection{Generalising probabilities $h(p)$ 
and $g(p)$ when there are deceptive agents}
 To evaluate the expected utility of a specific trustworthy 
 agent amidst deceptive agents, we consider the case where the total 
 of $H$ agents is divided into $H_t$ trustworthy 
 agents (who stick to probabilistic sniping) 
 and $H_d$ deceptive agents who pretend to restrict 
 their sniping but in reality enter the sniping race 
 at every possible occasion (obiously, $H_t+H_d = H$). 
 We can now  build on the earlier exposition 
 in section \ref{sct:util_prob_sniping}
 where, in eqs.~(\ref{eq:A_and_B_fion_of_p_and_gamma}) and 
 (\ref{eq:C_and_D_fion_of_p_and_gamma}),  we expressed the expected utilities 
 in terms of the probabilities $h(p)$ and $g(p)$. 
 Closer inspection of these equations shows that only 
 these probability functions need to be modified in the presence 
 of deceptive agents as the other factors only 
 depend on the game parameters. 
 We therefore introduce the following 
 notation which extends the  probabilities 
 $h(p)$ and $g(p)$:  
 
 \begin{itemize}
     \item $h_{tm}(p, H_t, H_d)$ is the probability that a  
     specific individual {\bf trustworthy} agent, who 
     is acting as {\bf market maker}, will {\bf lose} the 
     MZ-race given that the $H_d$ deceptive agents 
     snipe for sure, while the remaining $H_t-1$ trustworthy agents enter the sniping race with probability $p$
     (denoted as: wp. $p$):  

 \begin{eqnarray}
     h_{tm}(p,H_t,H_d) := P\left\{ \mbox{MM will {\bf lose} race}
     \given  \mbox{$H_d$ agents snipe for sure}, \ldots
     \right.\nonumber\\ 
   \left.\mbox{\ldots while $(H_t-1)$ remaining trustworthy agents snipe w.p.\! $p$}
     \right\}
     \label{eq:h_tm}
 \end{eqnarray}
 In addition to the market maker, the 
 number of agents in the race is $H_d$ (all the deceptive agents snipe for sure) plus  
 a sample (of stochastic size $N_t \sim Bin(H_t-1,p)$) 
 of the remaining $H_t-1$ trustworthy 
 agents who snipe with probability $p$. 
 Hence: 
 \begin{equation}
  h_{tm}(p, H_t, H_d) = \E \left( \frac{H_d + N_t}{1+H_d + N_t} \right)
 \quad \quad \mbox{where} 
 \quad \quad N_t \sim Bin(H_t-1,p)
\label{eq:h_tm_orig}
 \end{equation}

 \item $g_{tb}(p, H_t, H_d)$ is the conditional probability 
 that a specific {\bf trustworthy agent}, acting as 
 {\bf bandit}, {\bf wins} the sniping race, given that
 {\it he actually enters the race}, and 
  $H_d$ deceptive agents snipe for sure, while the remaining $H_t-1$ trustworthy agents enter the  race with probability $p$:

 \begin{eqnarray}
     g_{tb}(p, H_t, H_d) &:=& P\left\{ \mbox{agent will {\it win race}}
     \given  \mbox{ he enters the race as bandit,  } \ldots \right.\nonumber\\ 
     & &\quad \ldots \mbox{while the remaining $H_t-1$ trustworthy agents snipe w.p. $p$, }\ldots \nonumber  \\
    & & \left.\quad \ldots \mbox{and the $H_d$ 
    deceptive agents snipe for sure}
     \right\} 
     \label{eq:g_tb}
 \end{eqnarray}

A similar reasoning implies that $g_{tb}(p)$ 
 is a weighted 
 average of the following two terms: 
 \begin{itemize}
     \item if the market maker happens 
     to be selected from the group of 
     trustworthy agents (probability $(H_t-1)/(H-1)$) then 
     the list of agents in the race is composed of: 
     the agent under scrutiny, the market maker, 
     all the deceptive agents ($H_d$) and a random 
     selection of the remaining 
     $H_t-2$ trustworthy agents; 
     
     \item if the market maker happens to be selected 
     from the group of deceptive agents (probability 
     $H_d/(H-1)$), the list of 
     agents in the race comprises, in addition to 
     the trustworthy agent under scrutiny, the market 
     maker, all the remaining deceptive agents 
     ($H_d-1$), as well as a random selection from 
     the remaining $H_t-1$ trustworthy agents. 
     
 \end{itemize}
 Recasting this logic in mathematical parlance, we 
 arrive at the following expression: 
 
 \begin{eqnarray}
     g_{tb}(p, H_t, H_d) & =& \left(\frac{H_t-1}{H-1}\right)
     \E \left(\frac{1}{2+H_d + N_t^{(m)}} \right) 
      + 
     \left(\frac{H_d}{H-1}\right)
      \E \left(\frac{1}{2+(H_d-1)+N_t^{(b)}} \right)  
   \nonumber \\
   & =& \left(\frac{H_t-1}{H-1}\right)
     \E \left(\frac{1}{2+H_d + N_t^{(m)}} \right) 
      + 
     \left(\frac{H_d}{H-1}\right)
      \E \left(\frac{1}{1+H_d+N_t^{(b)}} \right)  
   \nonumber \\
 \mbox{where} & &
       N_t^{(m)} \sim Bin(H_t-2;p)  
      \quad\quad \mbox{and} \quad \quad 
       N_t^{(b)} \sim Bin(H_t-1;p) 
        \label{eq:g_tb_orig} 
 \end{eqnarray}
 (For the sake of 
 completeness we add that the binomial random variables are 
 independent).
 \end{itemize}
 An alternative characterisation of $g_{tb}$ which, from a computational point of view, is slightly more convenient is presented in 
 appendix \ref{appx:alternative_compute_g_tb}. 
 
 \begin{verbatim}
   See: KPD_Supplementary_material_2.ipynb (section.3)
 \end{verbatim}

\subsubsection{Non-compliance detection 
based on sequential  utility testing}
\label{sct:sprt}

Rather than basing the non-compliance test on 
the aggregated utility (which is a rather crude 
measure), 
we monitor the rate of occurrence of each individual 
utility outcome. 
To this end, we  utilise the fact that since 
 the game parameters 
 are common knowledge, each agent can use 
 Table~\ref{tab:Pay-off_referring_to_table1.p1201} 
 to compute the values of the $J=9$ different possible 
 utilities.  
 In addition, combining  the event probabilities 
 in that table with the generalisation $h_{tm}$ 
 and $g_{tb}$ 
 specified in eqs.~(\ref{eq:h_tm_orig})   
 and  (\ref{eq:g_tb_orig}),  he will be able to 
 compute the corresponding probability 
 of each utility outcome, both under the assumption 
 of compliance ($H_d=0$) and non-compliance ($H_d>0$).  
 These probabilities are listed in 
 Table~\ref{tab:util_prob_overview_1}.
 
\begin{table}[htbp]
    \centering
    \begin{tabular}{r|c}
    utility & probability \\
    \hline
    & \\[2ex]
        0  & $  \frac{1}{H}
\left\{\beta(1-2\overline{\mu})(1-h_{tm}(p)) \right\} 
+ \frac{H-1}{H} \left\{ 1 - 
\beta p g_{tb}(p)(1-\overline{\mu}) 
\right\} $\\[2ex]
 $ -\gamma(2\sigma-s) $ &  $\overline{\alpha} \beta \cdot \left(\frac{1}{H}\right) \cdot h_{tm}$\\[2ex]
 $s$                   & $ \left(\frac{1}{H}\right)\left\{\overline{\alpha}\beta h_{tm}(p) + (1-\beta)(1-2\overline{\alpha} - \overline{\mu})\right\}$\\[2ex]
 $-\gamma(\sigma-s)$    & $\left(\frac{1}{H}\right) \left\{\overline{\alpha}(1-\beta)+\beta h_{tm}(p) (1-2(\overline{\alpha}+\overline{\mu}))+\beta \overline{\mu}\right\}$\\[2ex]
 $ s+\sigma$           & $\left(\frac{1}{H}\right) \left\{\beta \overline{\mu} (1-h_{tm}(p)+\overline{\alpha}(1-\beta)\right\}$\\[2ex]
 $ 2s$                 & $\overline{\mu} \cdot \left(\frac{1}{H}\right) \left\{\beta h_{tm}(p) + (1-\beta)\right\} $\\[2ex]
 $ 2 \sigma -s $       & $\frac{H-1}{H}  \left\{ \overline{\alpha} \beta p g_{tb}(p) \right\} $\\[2ex]
 $\sigma -s$           & $ \left(\frac{H-1}{H}\right)\cdot\beta p g_{tb}(p)  \left\{ 1-2\overline{\alpha} -\overline{\mu}\right\}  $\\[2ex]
 $- \gamma s$          &$  \left(\frac{H-1}{H}\right)\cdot \frac{\beta}{2} \cdot\overline{\alpha} p g_{tb}(p)$ \\[2ex]

    \end{tabular}
    \caption{Overview of unique utilities and 
    corresponding probabilities (for a trustworthy agent).}
    \label{tab:util_prob_overview_1}
\end{table}

\begin{figure}[htbp]
    \centering
    \includegraphics[width=.9\textwidth,height=7.5cm]{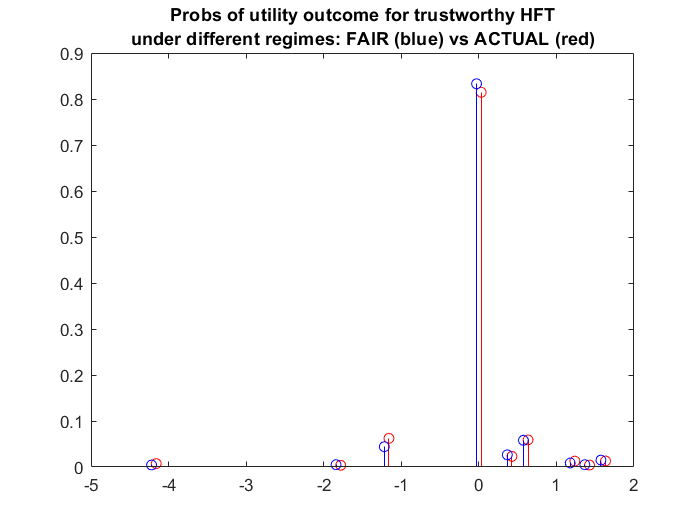}
     \includegraphics[width=.9\textwidth,height=7.5cm]{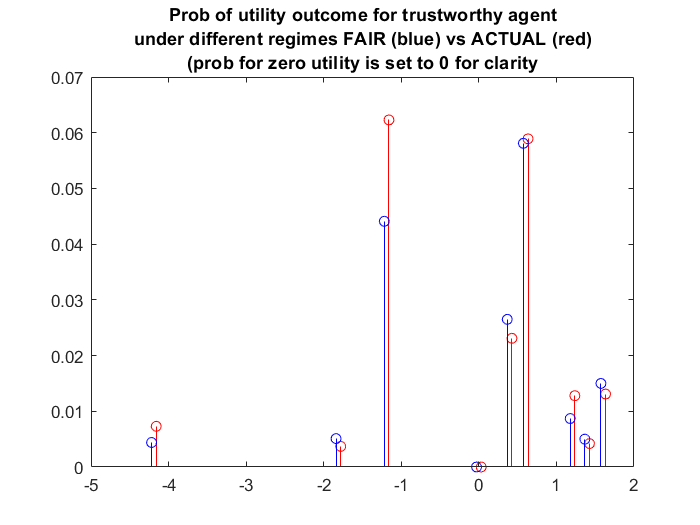}
    \caption{Probability distribution (taken 
    from Table~\ref{tab:util_prob_overview_1}) for the finite 
    set of utility outcomes under two different 
    regimes (BLUE: Fair ($H_d=0$),  RED: $H_d = 1$). Bottom: 
    Same figure as above, but probability of 
    zero utility has been suppressed for reasons of  
    clarity. 
    }
    \label{fig:stem_utility}
\end{figure}

 So, in essence, we have a situation in which each 
 stage game produces a utility outcome 
 $u_j \, (j = 1,2,\ldots, J)$ with a 
 known probability 
 distribution $\mathbf{\pi}^0 = (\pi^0_1, \ldots, \pi^0_J)$ under the 
 assumption of overall compliance (null-hypothesis of fair behaviour 
 with everyone sniping probabilistically at the optimal 
 rate $p^*_K$) 
 and an alternative known distribution 
 $\mathbf{\pi}^1 =  (\pi^1_1, \ldots, \pi^1_J)$  
 that applies to a trustworthy  agent, assuming that there are 
 a specific number $H_d > 0$ of  deceptive agents 
 that snipe for sure (see Appendix \ref{appx:prob_dist_unique_util}). 
This is illustrated in Fig~\ref{fig:stem_utility}  where we have plotted the 
probability of occurrence of the nine possible outcomes 
under two different regimes (with and without deceptive 
agents). 

When, after stage game $t$, an individual trustworthy agent 
receives utility $u_t$, he can compute the likelihood 
of this outcome under the two alternative hypotheses 
i.e. without (null hypothesis ${\cal H}_0$) or with 
(alternative hypothesis ${\cal H}_1$) deceptive snipers, 
and determine which one is more likely: 

$$  P(u_t \given {\cal H}_0) \leq ?? \geq P(u_t \given {\cal H}_1) $$

\begin{verbatim}
For illustrative purposes we look at a concrete example, 
See, KPD_supplementary_material_game_simulations.ipynb (section 9)
\end{verbatim}

Wald's sequential probability ratio test (SPRT) \cite{wald1945} 
provides a principled way of aggregating the evidence 
collected over a sequence $u_1,u_2, \ldots, u_t$ of 
utility observations.  More specifically,  at every 
time (stage game) $t$ we construct the following 
test statistic: 
\begin{equation}
S(t) = \sum_{\tau = 1}^t 
\log\frac{P(u_\tau \given {\cal H}_1)}{P(u_\tau \given {\cal H}_0)} .
\label{eq:sprt}
\end{equation}
To implement the  sequential test procedure we first 
need to compute the following two threshold values: 
\begin{equation}
 a :=  -\log \left(\frac{1-\alpha}{\beta}\right)  
\quad\quad b := \log \left(\frac{1-\beta}{\alpha} \right) 
\label{eq:sprt_thresholds}
\end{equation}
where $\alpha$ is type I error (related to significance) 
and $\beta$ is the type II error (related to power).  
We then  get the following stopping rule: 
\begin{itemize}
    \item if $a < S_t <b $: continue testing;
    \item if $S_t < a$:  accept ${\cal H}_0$ (all HFTs 
    are trustworthy)
    \item if $S_t >b $:  accept ${\cal H}_1$ (i.e. reject ${\cal H}_0$, 
    there is at least one devious HFT)
\end{itemize}

\begin{figure}
    \centering
    \includegraphics[height=5cm,width=\textwidth]{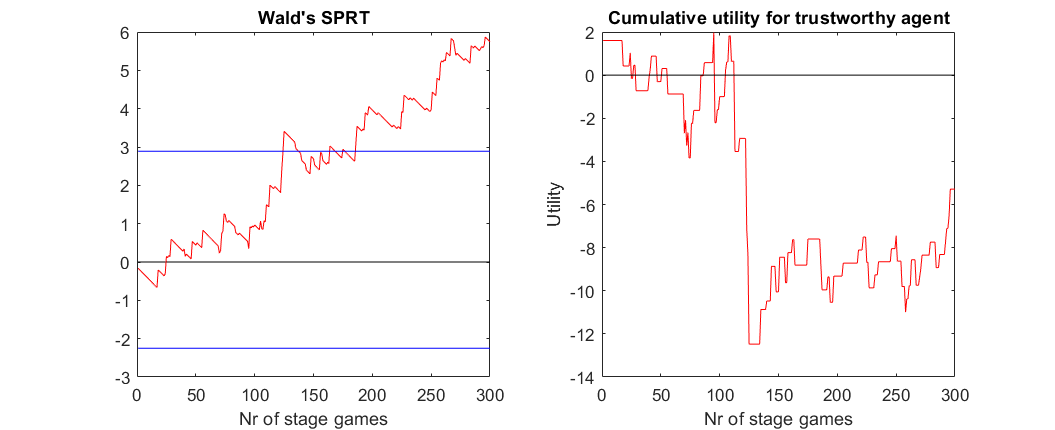}
      \includegraphics[height=5cm,width=\textwidth]{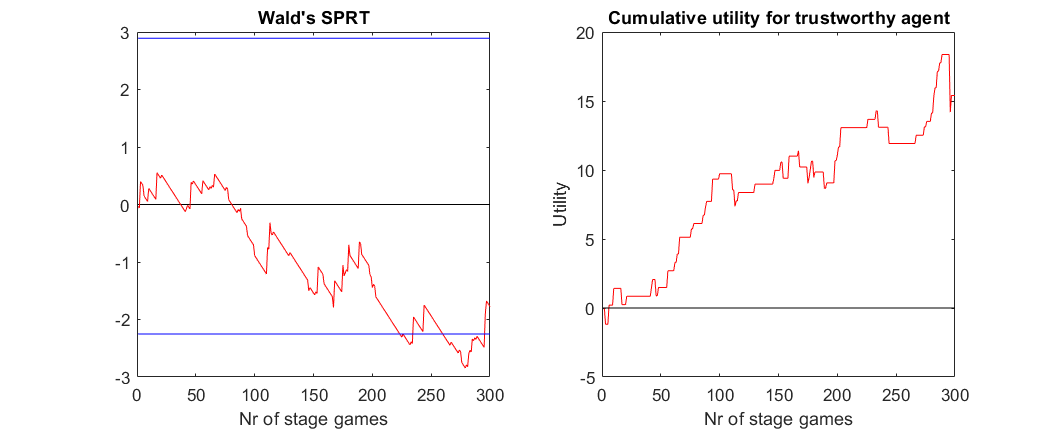}
    \caption{TOP: Simulation results 
    for a group of 4 agents, three of which are trustworthy and snipe with optimal probability $p^*_K$ while one of them is 
    deceptive and snipes for sure. The evolution of the utility of one 
    of the trustworthy agents during the repeated game  is plotted on the right, 
    and the corresponding Wald SPRT statistic on the left. Around stage game 115 the Wald statistic hits the upper limit ($b$ in eq.~\ref{eq:sprt_thresholds}) indicating that the null hypothesis of FAIR game can be rejected. 
    This would therefore be the moment the trustworthy agent would choose to quit the game. 
    BOTTOM: Same statistic and data but this time 
    for a group of four agents that are all abiding by implicit probabilistic sniping contract. In this 
    case the Wald statistic hits the lower limit 
    ($a$ in eq.~\ref{eq:sprt_thresholds}) around 
    stage game 175, allowing the agent to accept the null hypothesis of FAIR game. 
    }
    \label{fig:sprt}
\end{figure}

 So, any trustworthy agent that is monitoring his 
 accumulated utility will compose the Wald test-statistic 
 as follows: at time (stage game) $\tau$  he will 
 receive utility $u_{j(\tau)}$. Once this result 
 is known, he will compute 
 the probability ratio 
 $\pi^1_{j(\tau)}/\pi^0_{j(\tau)}$of this event under both 
 hypotheses (fair ${\cal H}_0$ and deceptive ${\cal H}_1$) and then 
 sum the logs of these ratios.  He will then 
 continue testing until hitting one of the two 
 thresholds in eq.(\ref{eq:sprt_thresholds}),   whereupon he can draw a statistically 
 significant conclusion regarding the compliance of 
 the other agents.  This is illustrated in 
 Fig.~\ref{fig:sprt} where the top row shows the 
fate of a trustworthy agent in a group of three additional 
agents one of which is cheating. 
The bottom row shows the fate of a trustworthy agent 
when each of his three colleagues is honouring the 
implicit contract of optimal probabilistic sniping.

\begin{verbatim}
See, KPD_supplementary_material_game_simulations.ipynb (section.10)
\end{verbatim}

\section{Related work}
\label{sct:related_work}

\paragraph{High-frequency trading }
Over the last decades, financial markets have changed significantly and become fragmented. Hence, traders can search across many markets, but 
this requires costly infrastructure and technology 
to make trading profitable \cite{Delaney2018}. In addition, high speed 
trading is now 
considered to be a crucial part of the 
trading technology. In 2010, the speed of round-trip between NASDAQ and the Chicago Mercantile Exchange decreased from over 14.5 milliseconds to under 8.1 milliseconds in 2014 \cite{Baldauf2014}. Also NASDAQ supports 
high frequency 
traders by offering faster access to their infrastructure and trade data transmission \cite{KAUFFMAN2015}.  
In response to these drastic changes in financial markets, 
{\it algorithmic trading} 
harnesses 
advanced technologies to connect to different markets directly and trade 
ever faster at lower costs. Therefore, 
high-frequency traders (HFTs)  
can trade continuously and benefit from serial order processing. This continuous time trading has given rise to a speed race and 
sniping has 
emerged 
as an opportunity 
for HFTs 
to benefit from a stale quote~\cite{Hagstromer2013}. In 
\cite{Budish2015} it is shown that HFTs will tend to post 
thin but strictly positive 
 bid-ask spreads to compensate the costs of getting sniped even without asymmetric information about fundamentals. 

The emergence of high-speed trading has spurred on a lot of research regarding the 
effects on important market parameters such as liquidity. 
The seminal paper by Menkveld and Zoican \cite{mz2017} has been the starting point 
for this research and has been discussed extensively in this paper. 
The focus in both this  and the MZ paper has been on a stylized version of 
the interactions occurring at high frequency electronic exchanges.  
However, real market are competitive and continuous, 
and algorithmic traders need to choose their strategies from scratch and for a longer period in order to benefit from the dynamic of the market and speed (arms race). Moreover, traders should have a set of predefined strategies \cite{Cason2019} and  traders' behaviour may be influenced by 
observing their competitors 
\cite{Romero2018}. 
Therefore, we explain agent's strategies and the reason why repeated game instead of one shot Nash equilibrium is a better choice to explain behaviours in the electric exchange. %
The case for repeated games is further strengthened by research by 
Bruce (2007) that shows that a one-shot stage game brings significant surplus loss to the traders in a dynamic setting \cite{Bruce2007}. More recently, some studies showed that repetition in the game allows decision makers (agents, players in the market) to construct their strategies explicitly and benefit in the long-run \cite{DalBo2017,Romero2018}.


Breitmoser(2015) analyzed individual strategies that are cooperate after mutual cooperation, defect after mutual defection, otherwise randomize. He finds a semi-grim equilibrium  as a threshold that agents start cooperating in the first round by knowing treatment parameters and switch to semi-Grim with equal probability in the repeated game. At the end, sustain mutual cooperation led to the long-run welfare \cite{Breitmoser2015} and \cite{Fudenberg2012}. Furthermore, agents gain experience from the history of the game and trust is built \cite{Kagel2016} and an indefinitely repeated games with high continuation probability helps agents to mix strategies with minimal restrictions on the type and lengths of pure strategies \cite{Romero2018}. While repeated game helps sustain cooperation, it is important to monitor the game in case of any defections. In 1947, Wald \cite{wald1945} develops a statistical test called Sequential probability ratio test (SPRT) based on likelihood ratio. This test is applied repeatedly during the sampling process and terminate whenever there is sufficient evidence in the data for one of the hypotheses of interest. In this paper, we use SPRT to detect non-compliance agent in different scenarios. 

\section{Conclusions and further research}
\label{sct:conclusions}

In this paper we re-analysed the MZ game, 
a stylized version of a market interacting with a  high frequency exchange as  originally described in \cite{mz2017},  
and argued that a recasting of the problem in terms of 
repeated games offers a natural interpretation 
in which the solution space of equilibria 
accommodates genuine probabilistic sniping. 
Probabilistic sniping was introduced in 
the original paper 
to account for changes in the behavior of HFTs  
when risk 
aversion increases, 
but its practical significance in 
the single-shot game is limited. 
However, {\bf in the context of 
repeated games} probabilistic sniping provides 
genuinely new and interesting insights.  
In particular: 

\begin{itemize}

\item We outlined a simple geometric argument 
that allows one to deduce the general 
conditions that govern the transitions 
in sniping behaviour. By translating these 
conditions in terms of the risk aversion 
parameter $\gamma$  we obtained  
two new thresholds 
($\overline\gamma_K$ and $\overline\gamma_L$).

\item Contrary to the 
    situation in the single-shot game, we 
    predict that probabilistic sniping will start playing a role even before  the 
    equilibrium utility for sure sniping 
    is reduced to zero (by increasing 
    risk aversion).

\item We defined unfavourable scenario in which there are deceptive agents in the game. In comparison  with favorable (fair) situation, the presence of non-compliant agents reduces trustworthy agent's  expected utility. We showed how one can use Wald's SPRT to  monitor the sequential game and flag when the utility for trustworthy agents deviates from the optimum point. 
\end{itemize}
 
\paragraph{Some notes on extensions 
for further research}

In both this paper and the original MZ paper it was shown how simple 
and highly stylized models can already provide unexpected insights 
and interesting conclusions.  We therefore expect that more realistic models 
will yield even more intriguing insights. Some of the challenges we intend 
to tackle in an upcoming paper are:

\begin{itemize}

    \item   We showed how important quantities that determine 
    the qualitative transitions in the results (such as 
    $s^*_K$ or $\overline\gamma_K$) can be computed based on the knowledge of the 
    game parameters, i.e. $\alpha, \mu, \delta$ 
    and $H$.  However, 
    in most realistic situations, 
    these parameters are unknown and 
    need to be estimated from observations.  As a result there is an amount of 
    uncertainty and noise associated with these results. 
    How stable are the conclusions with respect to 
    noisy game parameters? 

  \item In the repeated game scenario 
  expounded in this paper, each stage game 
  starts from 
  zero position. 
  It can be modified to a more realistic strategy, as the traders 
  want to regress back to zero-positions. 
  
  \item Risk aversion varies depending on position. One reason for this is 
  that traders dread reporting bad news. Hence, when they find themselves 
  in a negative position, they are willing to take more risk since, if they are lucky,  
  they can redeem themselves, and if they are unlucky, 
  the "size" of their bad news does not make 
  all that much difference. 
  
\end{itemize}

\paragraph{Acknowledgements}
The authors are very grateful to Profs. Menkveld and Zoican for their detailed feedback on earlier drafts of this paper.


\clearpage

\appendix

\begin{center}
    {\Large \bf Appendix: Supplementary Material }
\end{center}

\section{List of notation and abbreviations} 
\label{appx:abbrevs}

\begin{tabular}{|r|l|}
\hline
  MZ or MZ$^1$ & Menkveld-Zoican (single-shot) stage game\\
  MZ$^\infty$ &  infinite horizon repeated MZ game\\
  HFT   &  High Frequency Trader, also called {\it agent} \\
  LT    &  Liquidity trader (operates in background by 
  hitting available quotes)\\
  MM or M   &  market maker (selected from among the HFTs)\\
  B or HFB & (high frequency) bandit (all HFTs that were not selected as market maker)\\
 
  $H$  &  total number of HFT agents ($H = H_t+H_d$) \\
    $H_t$  &  total number of {\it trustworthy} agents (snipe probabilistically)\\
    $H_d$  &  total number of {\it deceptive} agents (snipe for sure)\\
  $v$ & intrinsic value of a (financial) asset\\
  $s$ &  (half) spread ($0\leq s \leq\sigma$), 
simply referred to as {\it spread}\\
  $p$ &  sniping probability \\
  $\alpha$ & arrival rate for news events (Poisson process)\\
  $\sigma$ & jump size upon news \\
  $\mu$ & arrival rate for LT arrivals (Poisson process) \\
  $\delta$ & exchange latency (duration of MZ race)\\ 
  $\gamma$ & risk aversion factor $(\gamma>1)$\\
  $\beta $ & probability that trigger  (i.e. first) event is news: $\beta = \alpha/(\alpha+\mu) $ \\
    $1-\beta $ & probability that trigger (i.e. first) event is LT 
    arrival: $ 1-\beta = \mu/(\alpha+\mu) $ \\
 $\overline{\alpha}$ &  (half) the expected number of news arrivals in interval $[0, \, \delta)$ \\
 $\overline{\mu}$ &(half) the expected number of LT arrivals in interval $[0, \, \delta)$ \\
 $m$ & $1-\overline\mu$ \\
 $\overline{\theta}$ &harmonic mean of $\overline{\alpha}$ 
 and $\overline{\mu}$  \\
  $s^*, u^*$ & spread and utility at point of indifference \\
  $s^*(p), u^*(p)$ &  same as above under probabilistic sniping\\
  $p^*_K$ & optimal sniping probability (maximises expected utility) \\
    $s_K^*$ &  corresponding optimal spread $s_K^* = s^*(p^*_K)$ \\
   $\overline{\gamma}_K$& $\gamma$-threshold for 
   transition from {\it sure} to {\it probabilistic} sniping \\ 
   $\overline{\gamma}_L$& $\gamma$-threshold for 
   transition from {\it probabilistic} to {\it no }sniping \\ 
   $h(p)$ & probability that market maker will {\bf lose} race (when HFTs snipe wp. $p$\\
   $g(p)$ & probability that bandit will win the race, given he enters race and all HFTs snipe wp. $p$\\
   $h_{tm}(p,H_t,H_d)$ & $h(p)$ for trustworthy market maker when 
   there are $H_t$ trustworthy and $H_d$ devious HFTs\\
   $g_{tb}(p,H_t,H_d)$ & $g(p)$ for trustworthy bandit when 
   there are $H_t$ trustworthy and $H_d$ devious HFTs \\
   
 \hline 
\end{tabular}

\clearpage
\section{Computer code} 
\begin{enumerate}

 \item \begin{verbatim}
     Simulation experiments:
     KPD_supplementary_material_game_simulations.ipynb
 \end{verbatim}
 \item \begin{verbatim}
     Check the existence of the solutions:
     KPD_supplementary_material_1.ipynb
 \end{verbatim}
\item \begin{verbatim}
     Check the sniping probability properties and numerical test:
     KPD_supplementary_material_2.ipynb
 \end{verbatim}
 \item \begin{verbatim}
     Check the numerical analysis:
     KPD_supplementary_material_numeric.ipynb
 \end{verbatim}

\end{enumerate}

\section{Details of payoff computation}
\label{sct:payoff_details}
The payoff for each agent is based on 
eq.~\eqref{eq:payoff_computation_1}  which we here restate: 
  \begin{align*}
    \mathbf{payoff}  &=  \mathbf{position}_{(t=\delta)} \times \mathbf{value }_{(t=\delta)} \,\, + \,\, \mathbf{income}_{(t=\delta)} &
\end{align*}
This allows us to compute the payoff for each event-code. See table (\ref{tbl:detailed_payoff_race}) and (\ref{tbl:detailed_payoff_no_race}) for detailed payoff and utility computation.

\begin{table}[htbp]
    \centering  
    \begin{adjustbox}{totalheight=\textheight-2\baselineskip}
  \begin{tabular}{|c|c|c|c|c|c|}
  \hline

  \multicolumn{1}{|c|}{Event codes}  & Values at $t=\delta$ &\multicolumn{2}{c|}{MM loses race } &\multicolumn{2}{|c|}{MM wins race }  \\  \cline{3-6}
   
  \multicolumn{1}{|c|}{ when there is a race} & & MM & B & MM & B\\ \hline

    \multirow{4}{*}{NG-NG} & Position =  & $-1$ & $+1$ & $0$ & $0$\\
    
    &  Value =   & $2\sigma$ &  $2\sigma$ &  $2\sigma$ &  $2\sigma$   \\ 
    
    &  Income =  & $s$ & $-s$ & $0$ & $0$   \\
    
    & payoff = & $-2\sigma+s$ & $2\sigma-s$ & $0$  & $0$ \\   \cline{2-6}
    
    & Utility = & -$\gamma(2 \sigma - s)$  & $ 2\sigma- s$  & $0$ & $0$ \\  \hline \hline

    \multirow{4}{*}{NG-NB} & Position =  &  $-1 $ &  $+1$ & $0$ & $0$\\
    
    &  Value =   & $0$ & $0$ & $0$ &  $0$  \\ 
    
    &  Income =  & $s$ & $-s$ & $0$ & $0$   \\
    
    & payoff = &  $0 +s$  & $0 -s$ & $0$  & $0$ \\  \cline{2-6}
    
    & Utility = &  $s$ &  -$\gamma s$ & $0$ & $0$ \\  \hline \hline

    \multirow{4}{*}{NG-LA} & Position =  &  $-1 $ & $0$ &  $ $-1 $ $ & $0$\\
    
    &  Value =   & $\sigma$ & $\sigma$ & $\sigma$ &  $\sigma$  \\ 
    
    &  Income =  & $s$ & $0$ & $s$ & $0$   \\
    
    & payoff = & $-(\sigma -s)$ & $0$ &  $-(\sigma -s) $ & $0$ \\  \cline{2-6}
    
    & Utility = &  -$\gamma(\sigma - s)$  & $0$ &  -$\gamma(\sigma - s)$  & $0$ \\  \hline \hline

     \multirow{4}{*}{NG-LB} & Position =  & $0$ &  $+1$ & $+1$ & $0$\\
    
    &  Value =   & $\sigma$ & $\sigma$ & $\sigma$ &  $\sigma$   \\ 
    
    &  Income =  & $2s$ & $-s$ & $-(-s)$ & $0$   \\
    
    & payoff = &$ (0) (\sigma)+ 2s $ & $\sigma -s$ & $\sigma + s$  & $0$ \\  \cline{2-6}
    
    & Utility = &  $2s$ & $\sigma -s$ & $\sigma + s $& $0$ \\  \hline \hline

     \multirow{4}{*}{NG-no} & Position =  &  $-1 $ &  $+1$ & $0$ & $0$\\
    
    &  Value =   & $\sigma$ & $\sigma$ & $\sigma$ &  $\sigma$   \\ 
    
    &  Income =  & $s$ & $-s$ & $0$ & $0$   \\
    
    & payoff = & $-(\sigma-s)$  & $\sigma -s$ & $0$  & $0$ \\  \cline{2-6}
    
    & Utility = &  -$\gamma (\sigma - s)$ & $\sigma-s$  & $0$ & $0$ \\  \hline \hline

    \multirow{4}{*}{NB-NG} & Position =  & $+1$ &  $-1$ & $0$ & $0$\\
    
    &  Value =   & $0$ & $0$ & $0$ &  $0$  \\ 
    
    &  Income =  & $s$ & $-s$ & $0$ & $0$   \\
    
    & payoff = &  $0 +s$  & $0 -s$ & $0$  & $0$ \\  \cline{2-6}
    
    & Utility = &  $s$ &  -$\gamma \, s $ & $0$ & $0$ \\  \hline \hline
      
      \multirow{4}{*}{NB-NB} & Position =  & $+1$ &   $-1 $ & $0$ & $0$\\
    
    &  Value =   &  $2\sigma$ &  $2\sigma$ &  $2\sigma$ &   $2\sigma$   \\ 
    
    &  Income =  & $s$ & $-2s$ & $0$ & $0$   \\
    
    & payoff = & $-2 \sigma +s$ &$ 2 \sigma -s$ & $0$  & $0$ \\   \cline{2-6}
    
    & Utility = & -$\gamma (2 \sigma - s) $&$ 2 \sigma - s$ & $0$ & $0$ \\  \hline \hline

    \multirow{4}{*}{NB-LA} & Position =  & $0$ &  $-1 $ &  $-1 $ & $0$\\
    
    &  Value =   & -$\sigma$ & -$\sigma$ & -$\sigma$ &  -$\sigma$  \\ 
    &  Income =  & $2s$ & $-s $& $s$ & $0$   \\
    
    & payoff = & $(0) (-\sigma) +2s$ & $0$ & $\sigma + s$  & $0$ \\  \cline{2-6}
    & Utility = &  $2s$  & $\sigma -s $&  $\sigma + s$ & $0$ \\  \hline \hline
    
    \multirow{4}{*}{NB-LB} & Position =  & +1 &  $0$ & +1 & $0$\\
    
    &  Value =   & $-\sigma$ &$ -\sigma$ & $-\sigma$ &  $-\sigma$   \\ 
    &  Income =  & $s$ & $0$ & $s$& $0$   \\
    
    & payoff = & -$\sigma+s$  & $0$ & -$\sigma+ s$   & $0$ \\  \cline{2-6}
    
    & Utility = & -$\gamma(\sigma - s)$ & $0$ & -$\gamma (\sigma - s)$ & $0$ \\  \hline \hline
    
      \multirow{4}{*}{NB-no} & Position =  & +1 &   $-1 $ & $0$ & $0$\\
    
    &  Value =   & -$\sigma$ & -$\sigma$ & -$\sigma$ &  -$\sigma$   \\ 
    
    &  Income =  & $-s$ & $s$ & $0$ & $0$   \\
    
    & payoff = & -($\sigma-s) $ & $\sigma-s $ & $0$  & $0$ \\  \cline{2-6}
    
    & Utility = &  -$\gamma(\sigma - s)$ & $\sigma -s$ & $0$ & $0$ \\  \hline \hline

\end{tabular}
\end{adjustbox}
\caption{This table represents a list of possible events, when there is a race, from $t = 0$ to $t=\delta$ with detailed  computation of payoff and utility based on eq.~\eqref{eq:payoff_computation_1}}
    \label{tbl:detailed_payoff_race}
\end{table}

\newpage

\begin{table}[htbp]
    \centering  
     \begin{tabular}{|c|r|c|c||c|c|c|c|}
      \hline

      \multicolumn{1}{|c}{Event code}  &  &\multicolumn{2}{c|}{No race } &\multicolumn{1}{c}{Event code}  &  &\multicolumn{2}{c|}{No race } \\ \cline{3-4} \cline{7-8}
   
  \multicolumn{1}{|c}{} & & MM & B &\multicolumn{1}{|c}{} & & MM & B \\ \hline

    \multirow{4}{*}{LA-NG} & Position =  &  $-1 $ &  $0$ &\multirow{4}{*}{LB-NG} & Position =  & +1 &  $0$ \\
    
    &  Value =   & $\sigma$ & $\sigma$  &  &  Value =   & $\sigma$ & $\sigma$\\
    
    &  Income =  &  $s $ & $0$      &   &  Income =  &  $s $ & $0$ \\
    
    & payoff = & -$\sigma + s$ &$0$   &  & payoff = &  $\sigma + s$ &$0$ \\   \cline{2-8}
    
    & Utility = & -$\gamma (\sigma - s)$ & $0$     & & Utility = &  $\sigma + s$ & $0$ \\  \hline \hline
  
    \multirow{4}{*}{LA-NB} & Position =  &  $-1 $ &  $0$ & \multirow{4}{*}{LB-NB} & Position =  & +1 &  $0$ \\

    &  Value =   & -$\sigma$ & -$\sigma$    & & Value =   & -$\sigma$ & -$\sigma$  \\ 
    
    &  Income =  &  $s $ & $0$  && Income =  & $ s $ & $0$  \\
    
    & payoff = & $\sigma + s$ &$0$ &&  payoff = &  -$\sigma + s$ &$0$ \\ \cline{2-8}
    
    & Utility = & $\sigma +s$ & $0$ && Utility = &  -$\gamma (\sigma - s) $ & $0$ \\  \hline \hline
   
    \multirow{4}{*}{LA-LA} & Position =  &  $-1 $ &  $0$ & \multirow{4}{*}{LB-LA} & Position =  & $+1$ &  $0$ \\
   
    &  Value =   & $0$ & $0$   && Value =   & $0$ & $0$  \\ 
    
    &  Income =  & s & $0$  && Income =  & $2s$ & $0$  \\
    
    & payoff = &  $s $ &$0$ && payoff = &  $2s $ &$0$ \\   \cline{2-8}
    
    & Utility = &  $s $ & $0$ &&Utility = &   $2s $ & $0$ \\  \hline \hline
  
    \multirow{4}{*}{LA-LB} & Position =  & $0$ &  $0$ & \multirow{4}{*}{LB-LB} & Position =  & +1 &  $0$ \\
   
    &  Value =   & $0$ & $0$   && Value =   & $0$ & $0$  \\ 
    
    &  Income =  &  $2s $ & $0$  && Income =  & $ s $ & $0$  \\
    
    & payoff = & $ 2s $ &$0$ && payoff = &  $s  $&$0$ \\   \cline{2-8}
    
    & Utility = &  $2s $ & $0$ &&Utility = &   $s $ & $0$ \\  \hline \hline

    \multirow{4}{*}{LA-no} & Position =  &  $-1 $ &  $0$ & \multirow{4}{*}{LB-no} & Position =  & $+1$ &  $0$ \\
    
    &  Value =   & $0$ & $0$   && Value =   & $0$ & $0$ \\ 
    &  Income =  & $s$ & $0$  &&  Income =  & $s$ & $0$  \\
    & payoff = & $s$ &$0$  && payoff = &  $s$ &$0$ \\   \cline{2-8}
    
    & Utility = & $s$ & $0$ &&  Utility = & $s$ & $0$ \\ \hline

\end{tabular}
\caption{This table represents a list of possible events, when there is no race, from $t = 0$ to $t=\delta$ with detailed  computation of payoff and utility based on 
eq.~\eqref{eq:payoff_computation_1}}
\label{tbl:detailed_payoff_no_race}
\end{table}
\clearpage
\section{Expected utility for market maker and bandit}
\label{appx:expansion_expected_util}

\begin{figure}[H]
    \centering
    \includegraphics[scale=0.7]{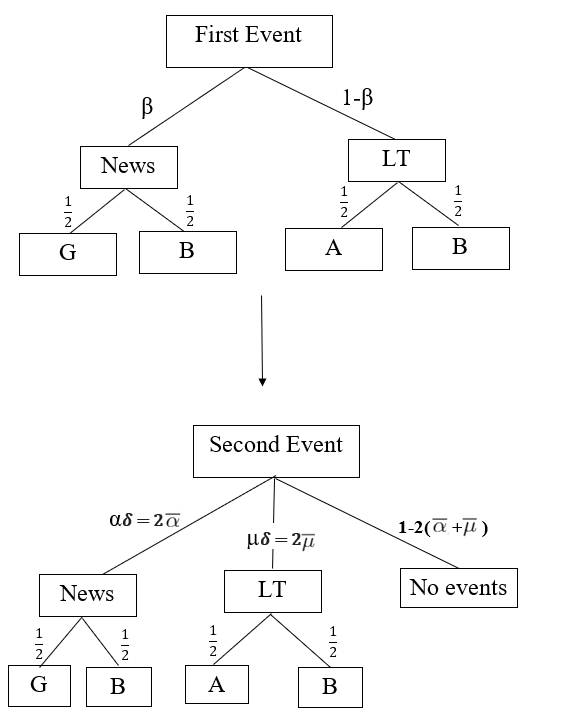}
    \caption{Schematic representation of the probabilities of the first 
    and second event. Since these two events are independent, the total 
    probability of any event code is the product of two of these probabilities. }
    \label{fig:event_probs_schematic}
\end{figure}

\paragraph{Expected utility for market maker}   
\quad\quad To compute the expected value $\E U_M$
we first condition on the nature of the 
trigger event: news (prob = $\beta$) or 
LT arrival (prob = $1-\beta$). 
Next, in case the trigger event is news,  
 a race ensues, and we 
additionally condition on whether (prob: $1-h$)
or not (prob. $h$) the market maker wins. 
To streamline the notation we split 
an eventcode $e$ 
into the first and 
second event $e = (e_1,e_2)$ (e.g. for 
$e= $ {\tt LA-NG } we have, $e_1 = $ {\tt LA } and $e_2= $ {\tt NG}. 
Referring to the probability tree in 
Fig.~\ref{fig:event_probs_schematic})   we see that the probability of 
each eventcode $e$ can be decomposed as: 

$$  
p(e) = p(e_1) p(e_2) = 
\left\{\begin{array}{lcl}
{\displaystyle \left(\beta\cdot \frac{1}{2}\right) \cdot p(e_2)} & \quad & \mbox{if }  e \in \mbox{{\tt N***} }\\[2ex]
{\displaystyle \left((1-\beta)\cdot \frac{1}{2}\right) \cdot p(e_2)} & \quad & \mbox{if }  e \in \mbox{{\tt L***} }
\end{array}
\right.
$$
Hence, we get the following 
expansion in terms of possible event-codes $e$: 
\begin{eqnarray}
\E U_M(s) &= &  \beta 
  \underbrace{\left[ \sum_{e \in N***} p(e) \,u_M(s \given e ) \right]}_{\mbox{Trigger = News (race!)}}  \quad 
 +  \quad (1-\beta)
\underbrace{\left[\sum_{e \in L***} p(e) \,u_M(s \given e)  \right]}_{\mbox{Trigger = LT  (no race!)}} \nonumber \\
 &= &  \beta 
 \left[  \frac{1}{2} \sum_{e_1\in N*}
 \sum_{e_2 } p(e_2) \,u_M(s \given e ) \right] \quad 
 +  \quad (1-\beta)
  \left[  \frac{1}{2} \sum_{e_1\in L*}
 \sum_{e_2 } p(e_2) \,u_M(s \given e ) \right]
\nonumber \\
&=&
  \frac{\beta}{2} 
   \sum_{e_1 \in N*} \sum_{e_2}
  \left[ \rule{0pt}{10pt} p(e_2)\,
  \left\{\rule{0pt}{8pt} h \cdot u_M(s \given e \, \& \,\mbox{MM loses}) 
+ (1-h) \cdot  u_M(s \given e \,\&\,\mbox{MM wins}) \right\}\right]  \nonumber\\[2ex]
&& \quad\quad\quad + \, 
\frac{(1-\beta)}{2} \sum_{e_1\in L*}\sum_{e_2 } \left\{p(e_2) \,u_M(s \given e)  \right\}  \nonumber
\label{eq:EU_M_schematically}
\end{eqnarray}

Based on the equation above, the following three sums need to be evaluated (using Table~\ref{tab:Pay-off_referring_to_table1.p1201}).  

\begin{enumerate}
    \item {\bf There is a race which MM loses:}

    \begin{eqnarray}
    S_1 &:=& \frac{1}{2}\sum_{e_1 \in N*} 
    \sum_{e_2} p(e_2)\,u_M(s\given e \, \& \, \mbox{MM loses})  \nonumber\\
    &= &
    \frac{1}{2}[
    2\overline{\alpha}\,s (1 - \gamma)\ -2 \,\overline{\mu}\,(\gamma(s-1) - 2s) + 2\,\gamma (s-1))]
    \end{eqnarray}

    \item  {\bf There is a race which  MM wins:} 
    \begin{eqnarray}
    S_2:= \frac{1}{2}\sum_{e_1 \in N*} 
    \sum_{e_2}
    p(e_2)\,u_M(s \given e_2\,\& \, \mbox{MM wins}) = \frac{1}{2}[2\overline{\mu}\, ((\gamma+1)s - \gamma +1) ]
    \end{eqnarray}

    \item {\bf No race:} 
    \begin{eqnarray}
     S_3 & := & \frac{1}{2} \sum_{e_1 \in L*} \sum_{e_2} p(e_2)\,u_M(s \given e) =  \frac{1}{2} \left[
    2\, \overline{\alpha}\, (s\,(\gamma-1)+1-\gamma)\, +2\,s(\overline{\mu}+1) \right]
    \end{eqnarray}

\end{enumerate}

Recombining we get: 

$$  \E U_M(s) = \beta \left( h S_1  +
(1-h) S_2 \right) +  (1-\beta)  S_3 
$$
In addition, it is useful to compute 
the values at the endpoints ($s=0$ and $s=1$):

\begin{equation}
\begin{array}{lcll}
C & = & \E U_M(s=0)  = &
\beta {h (-\gamma m)   + (1-h)(-\mubar(\gamma-1) } + (1-\beta)(\overline{\alpha} (1-\gamma))\\[1ex]
&&& = \beta h (\mubar q -m \gamma) - q\thetabar\\[2ex]
&&& = \beta h( 2\mubar q + \mubar -q-1) - q \thetabar\\[2ex]
D & = & \E U_M(s=1)  = & -\beta (\overline{\alpha} h q + m) + (\overline{\mu} + 1)
\end{array}
\end{equation}

\begin{verbatim}
   See: KPD_Supplementary_material_1.ipynb (section.2.3)
\end{verbatim}

\paragraph{Expected utility for bandit }

Since the bandit only gets utility when 
he wins the race, we have to condition 
three events: 

\begin{enumerate}
    \item there is a race: prob = $\beta$
    \item the bandit enters the race: prob = $p$
    \item he wins the race (given that he 
    entered): prob = $g(p)$  
\end{enumerate}
Using this information we can make the 
following factorisation: 

\begin{eqnarray}
  \E U_B(s) &=&\beta p g(p) \E U_B(s \given \mbox{bandit enters and wins race}) \\
  & = & \frac{1}{2}\beta p g(p) \sum_{e_1 \in N*} \sum_{e_2 } 
    p(e_2) \,  u_B(s \given  e_2 \,\&\, \mbox{bandit wins race})\nonumber\\
    &=& \beta p g(p) S
\end{eqnarray}
where we can use Table \ref{tab:Pay-off_referring_to_table1.p1201} to compute 
\begin{equation}
S := \frac{1}{2}\sum_{e_1 \in N*} \sum_{e_2} p(e_2)\,u_B(s\given e \,\&\,\mbox{bandit wins}) = 
\overline{\alpha}\,s (1-\gamma) + (\overline{\mu}(s-1)-s+1)
\end{equation}
From this we get: 

\begin{equation}
\begin{array}{lll}
A := \E U_B(s=0) &= (1-\mubar) \,\beta p g(p)  & =  m\, \beta p g(p)\\[2ex]
B := \E U_B(s=1) &= -\alphabar(\gamma-1) \, \beta p g(p) & =
-\alphabar q \, \beta p g(p)
\end{array}
\end{equation}

\begin{verbatim}
   See: KPD_Supplementary_material_1.ipynb (section.2.1)
\end{verbatim}

\clearpage
\section{Probability of sniping computation}
\label{sct:computation_h_and_g}

\subsection{Probability $h(p)$}

Recall from eq.(\ref{eq:def_hp})     
that $h(p)$ is defined as 
the probability that the market maker 
will {\bf lose} the race given 
that there is a race and all the 
bandits will attempt to snipe with probability 
$p$:
$$
 h(p):= P\left\{\mbox{MM will {\bf lose} MZ race} \given 
 \mbox{individual bandits enter race with prob $p$} \right\}
$$

It is conceptually easier to focus on the  complementary probability 
$\overline{h}(p) := 1-h(p)$ that the market maker will {\it win} the 
MZ-race and therefore thwart the sniping attempts of the bandits. Specifically, since the $H-1$
bandits decide independently and with equal probability $p$,  
whether or not they will snipe, the number of 
bandits that will enter the race is a random variable 
($N_B$ say) that is distributed according to 
a binomial distribution: 
$$ N_B \sim Bin(H-1,p).   $$
This is equivalent to saying that the 
probability of having 
$b=0,1,\ldots, H-1$ bandits participating in a race, equals: 
\begin{equation}
    P(N_B = b) = 
    \left(\begin{array}{c} H-1\\  b\end{array} \right)
    p^b (1-p)^{H-1-b} \quad  \mbox{for }  b = 0,1,\ldots, H-1
    \label{eq:bandit_in_race}
\end{equation}
When there are $b$ bandits racing, the probability that 
the market maker will win the race equals $1/(1+b)$ 
since each participant has the same probability 
of winning.  
As a consequence, the probability $\overline{h}(p)$ that 
market maker will win the race 
(and therefore escape sniping) 
equals
\begin{equation}
 \overline{h}(p) =  
\sum\limits_{b = 0}^{H-1} \frac{1}{b+1}
    \left(\begin{array}{c} H-1\\  b\end{array} \right)
p^b\,(1-p)^{H-1-b}   
\label{eq:MM_winning}
\end{equation} 
which can be recast as: 

\begin{equation}
\overline{h}(p) = 
\E\left(  \frac{1}{N_B +1}\right) \quad\quad 
\mbox{where}  \quad N_B \sim Bin(H-1,p).
\label{eq:hbar_expectation}
\end{equation}

\paragraph{Computation of $\overline{h}(p)$ and $h(p)$}  

To prove that eqs.~(\ref{eq:MM_winning}) and
(\ref{eq:hbar_expectation}) can be expanded to 
give rise to 

\begin{equation}
 \overline{h}(p) = \frac{1- (1-p)^H}{pH} , 
 \label{eq:prob_mm_escapes_sniping}
\end{equation}
we proceed as follows (we introduce $n=H-1$ 
for notational convenience). 

Recall that it is a straightforward consequence of the definition of binomial coefficients 
that: 
$$ \binom{n+1}{b+1} = \frac{n+1}{b+1} \binom{n}{b}  \quad \mbox{whence} \quad
\frac{1}{b+1}\binom{n}{b} = 
\frac{1}{n+1} \binom{n+1}{b+1} 
$$
Substituting this in eq.~(\ref{eq:MM_winning})
yields: 

\begin{eqnarray}
 \overline{h}(p) & = &
 \sum_{b=0} ^{n} \frac{1}{(b+1)}\binom{n}{b} p^{b} (1-p)^{n-b} \nonumber \\
 & = &
\frac{1}{p(n+1)} \sum_{b=0} ^{n} \binom{n+1}{b+1} p^{b+1} (1-p)^{(n+1)-(b+1)} \nonumber \\
& =&
 \frac{1}{p  (n+1)}\sum_{k=1}^{n+1} \binom{n+1}{k} p^k (1-p)^{(n+1) - k} 
 \quad\quad \mbox{(subs: $k = b+1$)}\nonumber\\
&=&
 \frac{1}{p (n+1)} \Bigg( 
 \underbrace{\sum_{k=0} ^{n+1} \binom{n+1}{k} p^k (1-p)^{(n+1-k)}}_{=1} - p^0 (1-p)^{n+1} \Bigg) \nonumber\\
&=&
\frac{1 - (1-p)^{n+1}}{p (n+1)} 
= \frac{1-(1-p)^H}{pH} \quad \quad (n = H-1).
\quad 
\nonumber
\end{eqnarray}
The computation above also implies  that 
\begin{equation}
    h(p) = 1-\overline{h}(p)= \E\left(\frac{N_B}{1+N_B}\right) =
    \frac{(1-p)^H -(1-Hp)}{Hp}
\label{eq;h_proved}    
\end{equation}
where  $N_B \sim Bin(H-1,p)$.

\paragraph{Some special cases}
Notice that for 
sure sniping (i.e. $p=1$) we get  
the familiar factor $h:=h(1) = (H-1)/H$.
Similarly, we can differentiate~eq.(\ref{eq;h_proved}) to obtain: %
\begin{equation}
    h'(p) = \frac{1-(1-p)^{H-1}(Hp+1-p)}{Hp^2} 
\quad \quad \Longrightarrow \quad\quad
h'(1) = \frac{1}{H}    
\end{equation}
To investigate the behaviour for small 
values $p \downarrow 0$  we use the 
standard binomial 
expansion: 
$$ (1-p)^H = 1 -Hp + \frac{H(H-1)}{2} p^2
-\frac{H(H-1)(H-2)}{6} p^3 + \ldots    $$
from which we conclude:   
\begin{equation}
h(p) = 
 \frac{H-1}{2}p - \frac{(H-1)(H-2)}{6}p^2 + o(p^3)  \quad \quad  \mbox{as } p\rightarrow 0. 
\label{eq:hp_expanded}
\end{equation}
and therefore: 
\begin{equation}
h(0) = 0 \quad \quad \mbox{and } \quad\quad h'(0) = \frac{H-1}{2}.
\label{h_dh_at_0}
\end{equation}

\subsection{Probability  $g(p)$}

Recall eq.(\ref{def:g(p)}) and (\ref{eq:g(p)})

\begin{equation}
g(p) := P\{ \mbox{individual bandit will  {\it win} race} \given  
 \mbox{he participates in the race} \}
 \nonumber
\end{equation}
Since the market maker always races, and we condition on this bandit also being 
in the race, the number of agents in the 
race equals $2+N_B$ where $N_B \sim Bin(H-2,p)$. 
Hence: 
$$ g(p) = \E\left(\frac{1}{2+N_B} \right)  $$
Using the binomial identity:
$$ \binom{n+2}{b+2} =  
\frac{(n+2)(n+1)}{(b+2)(b+1)} \binom{n}{b}  $$
a computation completely analogous to the 
one for $h(p)$ shows that:

\begin{equation}
g(p) =
\sum_{b=0} ^{H-2} \frac{1}{b+2}\binom{H-2}{b} p ^b (1-p)^{H-2-b}  = 
 \frac{(1-p)^H - (1 -pH)}{H(H-1)p^2} =  \frac{h(p)}{(H-1)p} 
\end{equation}

\paragraph{Special cases}

In the paper we need the following special cases: 

\begin{itemize}
    \item {\bf Limit for sure sniping $p\rightarrow 1$}
\begin{equation}
    g(1) = \frac{h(1)}{H-1} = \frac{1}{H}
\end{equation}
and also: 

\begin{equation}
    g'(p) = \frac{h'(p)p-h(p)}{(H-1)p^2}  \quad\quad 
    \Longrightarrow \quad\quad g'(1) = \frac{h'(1)-h(1)}{H-1} = 
    -\frac{H-2}{H(H-1)}
\end{equation}

\item {\bf Limit for no sniping $p \rightarrow 0$} \quad 
Using the expansion eq.~\ref{eq:hp_expanded} for $h(p)$, we get a similar expansion 
for $g(p)$:

\begin{equation}
    g(p) = \frac{1}{(H-1)p}
    \left\{ \frac{H-1}{2}p - \frac{(H-1)(H-2)}{6}p^2 + o(p^3)\right\}
    = \frac{1}{2} - \frac{H-2}{6}p + o(p^2)
    \label{eq:gp_expanded}
\end{equation}
From this we can conclude: 
\begin{equation}
    g(0) = \frac{1}{2} \quad \quad 
    \mbox{and} \quad\quad  g'(0) = - \frac{H-2}{6}. 
    \label{eq:g_and_dg_at_0}
\end{equation}

\end{itemize}

\begin{verbatim}
   See: KPD_Supplementary_material_2.ipynb
\end{verbatim}

\clearpage

\section{Proof of Proposition~\ref{thm:gamma_K}: Existence and 
uniqueness of \texorpdfstring{$\overline\gamma_K$}{Lg}}
\label{appx:trans_pure_prob}

To prove the existence 
and uniqueness of the transition threshold 
$\overline\gamma_K$  we need to show that 
the cubic polynomial (cf. eq.~\ref{eq:K_cubic}): 
$$ K(\gamma) = K_3\gamma^3 + K_2 \gamma^2 + 
K_1 \gamma +K_0   $$
has a unique zero-crossing greater than 1. 
To this end we prove the following three steps  
(also see 
Fig~\ref{fig:K_cubic_proof}): 

\begin{enumerate}
    \item $K_3 < 0$ indicating that $\lim_{\gamma\rightarrow \infty} K(\gamma) = -\infty$, i.e. $K(\gamma)$ will 
    be negative for sufficiently large values of  $\gamma$;
    \item $K(1) = K_3+K_2+K_1+K_0 > 0.$ which in 
    combination with $K_3 <0$,  shows that there is 
    at least one zero-crossing greater than 1; 
    \item Finally, we show  that 
    $K''(1) = 6K_3 + 2K_2 < 0$ 
    which proves that the cubic polynomial is concave to the right of  $\gamma=1$. This implies that the zero-crossing  $\overline{\gamma}_K$ is in fact unique.
\end{enumerate}
The detailed calculations that prove the 
three steps above can be found in the 
Python notebook: 
\begin{verbatim}
   See: KPD_Supplementary_material_1.ipynb (section.5)
\end{verbatim}

\begin{figure}[H]
    \centering
    \includegraphics[scale=0.6]{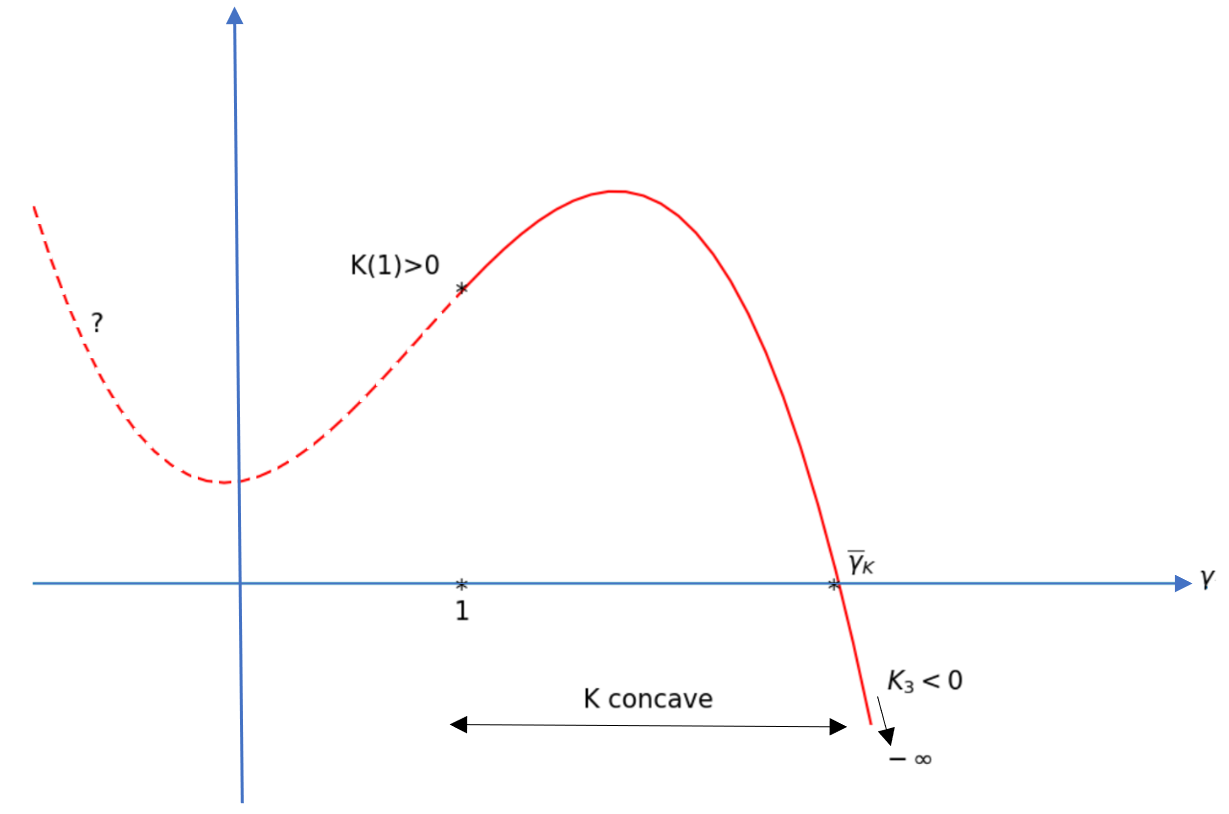}

    \caption{The proof of the existence and uniqueness of the threshold 
    $\overline\gamma_K$ is based on 
    the following three properties 
    of the $K(\gamma)$ cubic polynomial: (i) $K_3 < 0, K(1)>0,$ and $K$ is 
    concave to the right of 1. 
    }
    \label{fig:K_cubic_proof}
\end{figure}

\section{Proof of Proposition~\ref{thm:gamma_L}: Existence and 
uniqueness of \texorpdfstring{$\overline\gamma_L$}{Lg}}
\label{appx:trans_prob_non}

To prove that the transition threshold $\overline\gamma_L$ exists and is unique, we  point out 
that (cf. eq.~(\ref{eq:transition_prob_non_general_2}))
$$  
    \left. \frac{du^*(p)}{dp} \right|_{p=0} = 0
    \quad \Longleftrightarrow  \quad N'(0) = 0.
$$    
The right hand side can be expanded as a function 
of $\gamma$, or even more conveniently, as a function 
of $q  = \gamma-1$.  It turns out 
: 
$$
N'(0) =   -\frac{1}{2}\overline{\alpha}\beta\overline{\theta} q^2 + 
\frac{1}{2} \beta(1-\overline{\mu})(\beta\overline{\mu} - \beta + \overline{\mu} + 1). $$
Computing the zero-crossings for the 
quadratic equation on the right hand side, 
we conclude that there is a unique solution 
greater than $1$, viz.: 
$$ \overline\gamma_L =  
1 + \sqrt{\frac{(1-\overline{\mu}) Z}{\overline{\alpha}\, \overline{\theta} }} \quad \quad 
\mbox{where }\quad   Z = 1+\overline{\mu} - \beta(1-\overline{\mu}). $$
This concludes the proof. 

The detailed calculations that prove the 
three steps above can be found in the 
Python notebook: 
\begin{verbatim}
   KPD_Supplementary_material_1.ipynb (section.6)
\end{verbatim}

 \section{Alternative computation of \texorpdfstring{$g_{tb}$}{Lg} }
 \label{appx:alternative_compute_g_tb}
 Let $A$ be the stochastic variable representing the number of agents in the race.  Denote by $M \in {\cal T}\, ( {\cal D})$ 
 the event that the market maker is one of the trustworthy ($ {\cal T}$), or deceptive ($ {\cal D}$) agents, respectively. 
 Then 
 
 $$A = 
 (A \given M \in  {\cal T}) P(M \in  {\cal T}) + (A \given M\in   {\cal D}) P(M \in  {\cal D})   $$
We can make the following decomposition: 

\begin{itemize}
\item Since we condition on a specific trustworthy agent in the 
race, there are $H-1$ agents left, $H_t-1$ of them are trustworthy. 

\item We can visualise the selection process as follows.  Line up 
the agents.  The first $H_t$ are trustworthy, the next $H_d$ are 
the devious ones. Indicate who will race or not. 
Agent 1 will race (we condition on this event), 
from the remaining $H_t-1$ trustworthy agent a stochastic 
number $N_t \sim B(H_t-1,p) $ will race, as well as all the devious ones. 
Then select a market maker uniformly from agent 2 through $H$ 
(we have conditioned on the first agent to race as {\it bandit}). 

\begin{itemize}
    \item If the choice of market maker is among the devious 
    ones, nothing changes in the list of racing participants. 
    
    \item If the choice of market maker is one of the 
    $H_t-1$ remaining trustworthy agents there are two 
    possibilities.  If the agent was already singled out for 
    racing (prob = $N_t/((H_t-1))$), nothing changes. 
    
    If however, the agent was 
    not planning on racing 
    (prob = $(H_t-1-N_t)/(H_t-1)$), 
    he will now do so and the number of racers is 
    incremented by one to $1+N_t+H_d+1= 2+N_t+ H_d$. 
    Combining these observations we get: 
    
    $$ \left(\frac{N_t}{H_t-1}\right)
    \left( \frac{1}{1+N_t+H_d}\right) + 
    \left( \frac{H_t- 1 -N_t}{H_t-1}\right) 
    \left(\frac{1}{2+N_t+H_d}\right)$$
or again: 
 $$ \frac{1}{(H_t-1)} \left\{
    \frac{N_t}{1+N_t+H_d} + 
     \frac{H_t- 1 -N_t}{2+N_t+H_d}\right\}$$
    
\end{itemize}
    \item $P(M \in  {\cal T}) = (H_t-1)/(H-1)$ the market maker is 
    chosen among the remaining trustworthy 
    
    \item $P(M \in  {\cal D}) = H_d/(H-1)$
\end{itemize}

Combining all these together we get the following 
alternative for eq.(\ref{eq:g_tb_orig}): 

\begin{eqnarray}
g_{tb}(p) & = &
\left(\frac{1}{H-1} \right) \E \left\{
    \frac{N_t}{1+N_t+H_d} + 
     \frac{H_t- 1 -N_t}{2+N_t+H_d}\right\}  \nonumber\\
     && \quad + \left(\frac{H_d}{H-1} \right) \E \left\{
    \frac{1}{1+N_t+H_d} 
    \right\}  \nonumber \\
    &=& \left(\frac{1}{H-1} \right) \E \left\{
    \frac{N_t}{1+N_t+H_d} + 
     \frac{H_t- 1 -N_t}{2+N_t+H_d} 
      + 
    \frac{H_d}{1+N_t+H_d} 
    \right\} \\[2ex]
   && \quad \mbox{where } N_t \sim Bin(H_t-1,p).  \nonumber
\label{eq:g_tb_2}
\end{eqnarray}
 This characterisation of $g_{tb}$ has the advantage that 
 it involves only a single random variable, which makes 
 sense whenever $H_t\geq 1$.  The original 
 characterisation in eq.~(\ref{eq:g_tb_orig})  --- although conceptually simpler --- involves two binomial random 
 variables and 
 requires $H_t \geq 2$ to be well-defined..
 
 \begin{verbatim}
 See: KPD_supplementary_material_2.ipynb
 \end{verbatim}

\section{Change in mean utility due to deceptive agents}

 We are now in a position to 
 combine all these results and write down the 
 equivalents for eqs.~(\ref{eq:A_and_B_fion_of_p_and_gamma}) and 
 (\ref{eq:C_and_D_fion_of_p_and_gamma}).  
 More specifically, if a trustworthy agent acts 
 as bandit, his utility is a linear function 
 determined by the endpoints:

 \begin{equation}
\begin{array}{lcl}
 A_{tb}(p, H_t,H_d)  &=& m \beta p\, g_{tb}(p, H_t,H_d) \\[1.5ex]
  B_{tb}(p,H_t,H_d)  &=& -\overline{\alpha}(\gamma-1) 
  \beta p \,g_{tb}(p, H_t,H_d) \\[1.5ex]
  C_{tm}(p,H_t,H_d) & =  &
         -\left\{ q\overline{\theta}+\beta(m\gamma - \overline{\mu} q)  \,h_{tm}(p, H_t, H_d)\right\} \\[1.5ex]
     D_{tm}(p,H_t,H_d) & =  & 
        (1+\overline{\mu})- \beta \left\{m + \overline{\alpha} q \, 
        h_{tm}(p, H_t,H_d)\right\}
    \end{array}
    \label{eq:ABCD_fion_of_p_and_Hd}
\end{equation}
To evaluate for a given spread the 
expected 
utility $u_t(s)$ of a trustworthy HFT
one simply needs to evaluate the expected 
utilities in both roles (bandit and market maker): 

$$ u_{tb}(s) = A_{tb}(1-s) + B_{tb} \, s  \quad\quad 
\mbox{and} \quad\quad 
 u_{tm}(s) = C_{tm}(1-s) + D_{tm} \, s 
$$ 
which can then be combined (assuming that all agents 
advertise the same spread $s$ and therefore have 
equal probability of being selected as market maker): 
\begin{equation}
    u_t(s) = \frac{1}{H}\, u_{tm}(s) 
    + \frac{H-1}{H} \, u_{tb}(s) 
\label{eq:u_t}    
\end{equation}

\begin{verbatim}
See.KPD_supplementary_material_numeric.ipynb (section.6)
\end{verbatim}

\section{Probability distribution of unique utilities} 
\label{appx:prob_dist_unique_util}
In general, there are only 9 unique utilities (see table 1). 
This is still the case in the  presence of deceptive agents 
since even deceptive agents will publish the optimal spread. 
We can compute the probability with which these occur 
explicitly. 
Notice that there are three different cases: 

\begin{enumerate}
    \item Utilities that occur only for MM 
    ($-\gamma(2\sigma-s), s, -\gamma(\sigma-s), 2s,\sigma+s$)
    \item Utilities that occur only for bandit  
    ($2\sigma-s,-\gamma s, \sigma-s$)
    \item Zero utility that occurs for both market maker and bandit
\end{enumerate}

Let's do so for some concrete cases: 

\begin{itemize}
    \item {\bf Utility $2\sigma-s$ for successful sniper }
    (example of utility that occurs only for bandit): 
    this utility 
    occurs twice (NG-NG and NB-NB + MM loses race to successful 
    sniper). To compute the probability we need to multiply 
    the probability of each row ($(1/2)\overline{\alpha} \beta$) with that of the column and add it all together.  Since the column is the 
    same for both events (let's denote it as $p_{col}$ we get: 
    $$ p\{2\sigma-s\} = \overline{\alpha} \beta \cdot p_{col}$$
where 
$$
\begin{array}{rcl}
p_{col}  & = 
& P(\mbox{agent get this utility}\given \mbox{he's  bandit}) P(\mbox{he's bandit})  \\
& & \quad+ P(\mbox{agent get this utility}\given \mbox{he's  MM}) P(\mbox{he's MM})  \\[2ex]
 & = 
& P(\mbox{agent get this utility}\given \mbox{he's  bandit}) ((H-1)/H)  \\
& & \quad  + 0 \cdot (1/H)  \\[2ex]
 & = 
& {\displaystyle \frac{(H-1)}{H}}\, P(\mbox{agent get this utility}\given \mbox{he's  bandit})
\end{array}
$$
The last probability can be expanded further: 

$$  
\begin{array}{rl}
     &  P(\mbox{agent get this utility}\given \mbox{he's  bandit})  \\
    = & P(\mbox{agent enters race and wins} \given \mbox{he's  bandit})\\
= & P(\mbox{agent enters race } \given \mbox{he's  bandit}) \cdot
    P(\mbox{agent  wins race} \given \mbox{he's  bandit and enters race}) \\
=& p\, g_{tb}(p)    
\end{array}
$$
where we need to use $g_{tb}$ since we are only interested 
in the fate of trustworthy agents.  Combining both results we see 
that 
$$ P\{2\sigma - s\} =\overline{\alpha} \beta \cdot \left(\frac{H-1}{H}\right)
p\, g_{tb}(p) $$

\item {\bf Utility $\sigma-s$ for successful sniper }
    this utility 
    occurs 4 times when there is NG-NO, NB-NO and MM loses the race,and NG-LB and NB-LA and bandit benefits from the news arrival (MM loses the race in the news side) .
    
In all cases, the probability that successful sniper (bandit) enters the race $(H-1)/H$ and win the race is $ p\, g_{tb}(p)$ (As we explained in detail in the utility $P\{2\sigma - s\}$).

    \subsubitem In the case of NG-NO and NB-NO, the row probability is the same for both events.Combining with column probability (bandit is selected $\left(\frac{H-1}{H}\right)$ and is the race winner $p\, g_{tb}(p)$):
    $$ \mbox{prob}_{NG-No, NB-No} = \beta \cdot(1-2(\overline{\alpha}+\overline{\mu}))\cdot  \left(\frac{H-1}{H}\right)
p\, g_{tb}(p) $$
  \subitem and the same for NG-LB, NB-LA:

$$ \mbox{prob}_{NG-LB, NB-LA} = \beta \cdot\overline{\mu}  \cdot \left(\frac{H-1}{H}\right)
p\, g_{tb}(p) $$

The total probability of getting the utility ($\sigma -s$):
$$ P\{\sigma - s\} = \beta \cdot \left(\frac{H-1}{H}\right)
p\, g_{tb}(p) \cdot (1-2\overline{\alpha}-\overline{\mu} )   $$

\item {\bf Utility $-\gamma s$ for successful sniper } 
The only two events that produces this utility are NG-NB and NB-NG for 
a successful sniper (event probability 
in each case  $\overline{\alpha}\beta/2$ ). In this case, the 
trustworthy 
agent acts as a bandit (prob: $(H-1)/H$), 
he enters the race (prob: $p$) and wins 
(prob: $ g_{tb}(p)$). However, he doesn't benefit from it (make loss) as 
additional 
bad news   decreases the intrinsic value.  
Combining all this yields: 
$$ P\{- \gamma  s\} = 
\overline{\alpha} \beta \left(\frac{H-1}{H}\right)
p\, g_{tb}(p) $$

\item {\bf Utility $s$ for market maker. } This utility occurs on six 
occasions: NG-NB, NB-NG, LA-LA, LA-no, LB-LB and LB-no. 
For each of these cells in the table we can compute the probability 
by multiplying the probability of the row and the column. 
For the last four, the column probability is that of being market 
maker ($1/H$).  Adding the probabilities of the rows yields: 
$$ (1/2)(1-\beta)\cdot 2 \cdot (\overline{\mu}+1-2(\overline{\alpha}+\overline{\mu})) 
 = (1-\beta) (1-2\overline{\alpha} - \overline{\mu})
$$
Hence the total probability of the last four cells equals: 
$$\mbox{prob}_{LA-LA, LA-No, LB-LB, LB-No} = \frac{1}{H}  (1-\beta) (1-2\overline{\alpha} - \overline{\mu})   $$

To complete the computation we need to compute the probability 
of the first two cells. Only the probability of the column 
needs attention. It requires being selected as MM  (prob = $1/H$) 
but losing the race $h_{tm}(p)$.  Hence the combined total 
probability of these two cells equals: 

$$\mbox{prob}_{NG-NB, NB-NG} =  2\cdot (1/2) \overline{\alpha} \beta \cdot \left( \frac{1}{H}\right) 
h_{tm}(p) =  \left( \frac{1}{H}\right)\, \overline{\alpha}\, \beta 
h_{tm}(p).
$$
Combining the two results we get the following total 
probability for the utility $s$:

$$  
P\{s\} =  \left( \frac{1}{H}\right)\, \left[ \overline{\alpha}\, \beta 
h_{tm}(p) +  (1-\beta)(1 - 2\overline{\alpha} - \overline{\mu})\right]
$$

\item {\bf Utility $-\gamma(\sigma-s)$ for market maker. }

This utility occurs when there is: NG-LA, NB-LB, LA-NG, LB-NB, NG-No, and NB-No. 
The column probability for all cases is that of being selected as a market 
maker ($1/H$). 
Adding the probabilities of events in the rows, we see :

\subitem Probability for NG-LA and NB-LB:\\ 
In these events, MM wins and loses the race, so this utility happens four times and  $h_tm$ cancels out!.  

$$ \mbox{prob}_{NG-LA} = \left( \frac{1}{H}\right)\ 2 \cdot \frac{\beta}{2} \cdot \overline{\mu} \cdot h_{tm} + \left( \frac{1}{H}\right)\ 2 \cdot \frac{\beta}{2} \cdot \overline{\mu} \cdot (1-h_{tm})
$$
Therefor :
$$ \mbox{prob}_{NG-LA} = \left( \frac{1}{H}\right)\ \cdot \frac{\beta}{2} \cdot \overline{\mu}
$$
It is the same for NB-LB and combining two events:
$$ \mbox{prob}_{NG-LA,NB-LB} = \left( \frac{1}{H}\right)\ \cdot \beta \cdot \overline{\mu}
$$
\subitem Probability for NG-No and NB-No + MM loses the race with probability $h_{tm}(p)$: 
In both events, MM loses the race as bandit hits the stale quote:
$$ \mbox{prob}_{NG-No , NB-No} = \left( \frac{1}{H}\right)\ 2 \cdot \frac{\beta}{2} \cdot (1-2(\overline{\alpha}+\overline{\mu})) \cdot h_{tm}(p)
$$
\subitem Probability for LA-NG and LB-NB (The utility in these events happens four times): \\
$$ \mbox{prob}_{LA-NG , LB-NB} = \left( \frac{1}{H}\right)\ (1-\beta) \cdot \overline{\alpha} 
$$
Combining probabilities of six events:
$$P\{-\gamma(\sigma-s)\} =  \left( \frac{1}{H}\right)\ 
\left\{ \overline{\alpha} (1 - \beta)  + \beta h_{tm}  (1-2\overline{\alpha}-2\overline{\mu})+ \beta \overline{\mu} \right\}
$$
\item {\bf Utility $s + \sigma$ for market maker. }There are four corresponding events for this utility: NG-LB, NB-LA, LA-NB and LB-NG. 

To compute the probability:
\subitem To compute row probabilities for the first two events (NG-LB and NB-LA), we need to multiply the probability of each row: 

 $$ \mbox{prob}_{NG-LB, NB-LA} = \overline{\mu} \beta $$
 $$ \mbox{prob}_{LA-NB , LB-NG} = \overline{\alpha} (1-\beta) $$
 
 Then, to compute the column probability, we need the probability of($\frac{1}{H}$) MM being selected and ($1- h_{tm}(p)$) for winning the race (for NG-LB and NB-LA) 
 hence the total probability for this utility equals :
$$ P\{s + \sigma \} =\left(\frac{1}{H}\right) 
\left\{\beta \overline{\mu}  
(1- h_{tm}(p)) +\overline{\alpha} (1-\beta) \right\}$$

\item {\bf Utility $2s$ for market maker }
This utility befalls the 
market maker in four events : NG-LB, NB-LA, LA-LB and LB-LA. 

\subitem Row probability for NG-LB and NB-LA + MM loses the race:

    $$ \overline{\mu} \beta$$
Adding   $1/H$ the probability of being selected as MM and losing the race $h_{tm}$ yields:
     $$\mbox{prob}_{NG-LB, NB-LA} =  \overline{\mu} \beta \cdot \left(\frac{1}{H}\right) \cdot h_{tm}$$
\subitem  Row probability for  LA-LB and LB-LA: 
     $$ \mbox{prob}_{LA-LB, LB-LA} = \overline{\mu} (1-\beta) \cdot \left(\frac{1}{H}\right)$$
     
\subsubitem Adding the probabilities of events in the rows, we see :
$$ P\{2s \} =  \overline{\mu}  \left(\frac{1}{H}\right)
(\beta \cdot h_{tm}(p) + (1-\beta))$$

\item {\bf Utility $-\gamma(2\sigma-s)$ for market maker. } this utility occurs in events: NG-NG and NB-NB.

\subitem Adding row probability with column probability that is MM being selected ( $1/H$ ) and loses the race  $h_{tm}$ yields:
     $$P\{-\gamma(2\sigma-s) \} =  \overline{\alpha} \beta \cdot \left(\frac{1}{H}\right) \cdot h_{tm}$$

\item {\bf Zero utility $P\{0\}$ for bandit} 

This utility occurs in all events when the first trigger event is the arrival of LT. In these events, bandit has no chance of winning the race. Also, this utility occurs whenever bandit loses the race, either MM wins the race or the arrival of LT.

\subitem LA-LA, LA-LB, LB-LA, LB-LB:
      $$ \overline{\mu} (1-\beta) (2) \left(\frac{H-1}{H}\right)$$

\subitem LA-NG, LA-NB, LB-NG, LB-NB:
      $$ \overline{\alpha} (1-\beta) (2) \left(\frac{H-1}{H}\right)$$

\subitem LA-No and LB-No :
      $$ (1-2(\overline{\alpha}+\overline{\mu})) (1-\beta) \left(\frac{H-1}{H}\right)$$
\subitem NG-LA, NB-LB (bandit lose and win the race) :\\
    bandit loses the race : $$ P_{NG-LA} = \overline{\mu} \beta \left(\frac{H-1}{H}\right) \cdot p(1-g_{tb})$$
\\ bandit wins the race :  $$ P_{NG-LA} = \overline{\mu} \beta \left(\frac{H-1}{H}\right) \cdot p(g_{tb})$$
Therefore:
$$ P_{NG-LA} = \overline{\mu} \beta/2 \left(\frac{H-1}{H}\right) $$
\\ the same for NB-LB:
$$ P_{NB-LB} = \overline{\mu} \beta/2 \left(\frac{H-1}{H}\right) $$
\\ combining four events:
\\ $$ P_{NB-LB} = \overline{\mu} \beta \left(\frac{H-1}{H}\right) $$

\subitem When bandit loses the race in all other events if the first trigger event is news.\\
(events = NG-LB,NB-LA,NG-NG,NG-No,NB-No,NB-NG,NB-NB)

$$ P_{events} =\left(\frac{H-1}{H}\right) \beta (1-pg_{tb})(1-\overline{\mu}) $$

\subitem Total probability for $P\{0\}$ is: 
$$P\{0\}_{bandit} =  \left(\frac{H-1}{H}\right)(1-\beta+ \beta \overline{\mu}  + \beta (1-p g_{tb})(1-\overline{\mu})) $$

This can further be simplified to: 

$$P\{0\}_{bandit} =  \left(\frac{H-1}{H}\right)
(1-\beta p g_{tb}(1-\overline{\mu})) $$

\item {\bf Zero utility $P\{0\}$ for MM} 
\subitem e1 = NG-NG, NG-NB, NB-NG,NB-NB:
$$ P_{e1} = \left(\frac{1}{H}\right) 2 \beta \overline{\alpha} (1-h_{tm}) $$
\subitem e2 = NG-No, NB-No:
$$ P_{e2} = \left(\frac{1}{H}\right) \beta (1-2(\overline{\alpha} + \overline{\mu})) (1-h_{tm}) $$

Therefore :
$$ P\{0\}_{MM} = \left(\frac{1}{H}\right) \beta (1-h_{tm})(1-2 \overline{\mu})  $$

\end{itemize}

\paragraph{Alternative derivation}

$$P_B\{u=0\}:= \mbox{Prob that trustworthy 
bandit receives zero utility $(u=0)$}  $$

\begin{eqnarray}
P_B\{u=0\}&=&P_B(u=0\given \mbox{race})P(\mbox{race}) 
+ P_B(u=0\given \mbox{no race})P(\mbox{no race}) \nonumber \\
&=&P_B(u=0\given \mbox{race}) \cdot \beta
+ 1\cdot (1-\beta) 
\end{eqnarray}

\begin{eqnarray}
P_B(u=0\given \mbox{race}) &= & 
P_B(u=0\given \mbox{race and B enters race }) 
P(\mbox{B enters race} \given \mbox{race})
\end{eqnarray}

\clearpage

\begin{table}[]
    \centering
    \begin{tabular}{r|c}
    utility & probability \\
    \hline
    & \\[2ex]
        0  & $  \frac{1}{H}
\left\{\beta(1-2\overline{\mu})(1-h_{tm}(p)) \right\} 
+ \frac{H-1}{H} \left\{ 1 - 
\beta p g_{tb}(p)(1-\overline{\mu}) 
\right\} $\\[2ex]
 $ -\gamma(2\sigma-s) $ &  $\overline{\alpha} \beta \cdot \left(\frac{1}{H}\right) \cdot h_{tm}$\\[2ex]
 $s$                   & $ \left(\frac{1}{H}\right)\left\{\overline{\alpha}\beta h_{tm}(p) + (1-\beta)(1-2\overline{\alpha} - \overline{\mu})\right\}$\\[2ex]
 $-\gamma(\sigma-s)$    & $\left(\frac{1}{H}\right) \left\{\overline{\alpha}(1-\beta)+\beta h_{tm}(p) (1-2(\overline{\alpha}+\overline{\mu}))+\beta \overline{\mu}\right\}$\\[2ex]
 $ s+\sigma$           & $\left(\frac{1}{H}\right) \left\{\beta \overline{\mu} (1-h_{tm}(p)+\overline{\alpha}(1-\beta)\right\}$\\[2ex]
 $ 2s$                 & $\overline{\mu} \cdot \left(\frac{1}{H}\right) \left\{\beta h_{tm}(p) + (1-\beta)\right\} $\\[2ex]
 $ 2 \sigma -s $       & $\frac{H-1}{H}  \left\{ \overline{\alpha} \beta p g_{tb}(p) \right\} $\\[2ex]
 $\sigma -s$           & $ \left(\frac{H-1}{H}\right)\cdot\beta p g_{tb}(p)  \left\{ 1-2\overline{\alpha} -\overline{\mu}\right\}  $\\[2ex]
 $- \gamma s$          &$  \left(\frac{H-1}{H}\right)\cdot \frac{\beta}{2} \cdot\overline{\alpha} p g_{tb}(p)$ \\[2ex]

    \end{tabular}
    \caption{Overview of unique utilities and 
    corresponding probabilities.}
    \label{tab:util_prob_overview}
\end{table}

\end{document}